\documentclass[12pt]{article}
\usepackage{epsfig}
\usepackage{psfrag}
\usepackage{latexsym}
\usepackage{indentfirst}
\usepackage{fancyhdr}
\usepackage{amssymb}
\usepackage{amsmath}
\usepackage{amsfonts}
\usepackage{pifont}
\usepackage{cite}
\usepackage{color}
\usepackage{slashed}
\usepackage{graphics}
\usepackage{url}
\usepackage{array}
\usepackage{graphicx}
\usepackage{subfigure}

\def\be{\begin{equation}}
\def\ee{\end{equation}}
\def\bc{\begin{center}}
\def\ec{\end{center}}
\def\bea{\begin{eqnarray}}
\def\eea{\end{eqnarray}}
\newcommand{\bi}{\begin{itemize}}
\newcommand{\ei}{\end{itemize}}                 

\newcommand{\ba}{\begin{array}{c}}
\newcommand{\bad}{\begin{array}{ccc}}
\newcommand{\ea}{\end{array}}

\usepackage{color}
\definecolor{SR}{rgb}{0,0.2,0.8}
\definecolor{DM}{rgb}{0.5,0,0.5}
\definecolor{TO}{rgb}{1,0,1}

\allowdisplaybreaks 

\begin{document}

\begin{titlepage}
   \begin{center}
    {\Large\bf Renormalization Group Running of Fermion Observables in an Extended Non-Supersymmetric SO(10) Model}
   
   \vskip 0.2  cm
   \vskip 0.5  cm
Davide Meloni$^{\,a,}$\footnote{E-mail: \texttt{meloni@fis.uniroma3.it}},
Tommy Ohlsson$^{\,b,}$\footnote{E-mail: \texttt{tohlsson@kth.se}},
Stella Riad$^{\,b,}$\footnote{E-mail: \texttt{sriad@kth.se}}
\\[1mm]
   \vskip 0.7cm
 \end{center}

\centerline{$^{a}$ \it  Dipartimento di Matematica e Fisica, 
Universit\`{a} di Roma Tre,}
\centerline{\it  Via della Vasca Navale 84, 00146 Rome, Italy} 
\vspace*{0.2cm}
\centerline{$^{b}$ \it Department of Physics, School of Engineering Sciences,}
\centerline{\it KTH Royal Institute of Technology -- AlbaNova University Center,}
\centerline{\it Roslagstullsbacken 21, 106 91 Stockholm, Sweden}
\vspace*{1.5cm}

\begin{abstract}
\noindent
We investigate the renormalization group evolution of fermion masses, mixings and quartic scalar Higgs self-couplings in an {\em extended} non-supersymmetric SO(10) model, where the Higgs sector contains the ${\bf 10}_{\rm H}$, ${\bf 120}_{\rm H}$, and ${\bf 126}_{\rm H}$ representations. The group SO(10) is spontaneously broken at the GUT scale to the Pati--Salam group and subsequently to the Standard Model (SM) at an intermediate scale $M_{\rm I}$. We explicitly take into account the effects of the change of gauge groups in the evolution. In particular, we derive the renormalization group equations for the different Yukawa couplings. We find that the computed physical fermion observables can be successfully matched to the experimental measured values at the electroweak scale. Using the same Yukawa couplings at the GUT scale, the measured values of the fermion observables cannot be reproduced with a SM-like evolution, leading to differences in the numerical values up to around 80~\%. Furthermore, a similar evolution can be performed for a {\em minimal} SO(10) model, where the Higgs sector consists of the ${\bf 10}_{\rm H}$ and ${\bf 126}_{\rm H}$ representations only, showing an equally good potential to describe the low-energy fermion observables. Finally, for both the extended and the minimal SO(10) models, we present predictions for the three Dirac and Majorana CP-violating phases as well as three effective neutrino mass parameters.
\end{abstract}
\end{titlepage}

\section{Introduction}

There is a plethora of evidence for physics beyond the Standard Model of particle physics (SM). However, there is still no sign of any new physics at the Large Hadron Collider (LHC), which implies that the models for beyond-the-SM physics needs to be revised in order to fit the current constraints \cite{Aaboud:2016lwx,Aaboud:2016obm,Sirunyan:2016iap,Khachatryan:2016jqo}.

Among the numerous possibilities that one can imagine, grand unification theories (GUTs) are especially intriguing and are, in principle, motivated by the fact that the gauge couplings in the SM are close to unifying at some high-energy scale $M_{\rm GUT}$. In such theories, matter fields usually belong to one or more irreducible representations of the group and this allows to relate masses and mixings in a non-trivial way, reducing at the same time the number of the independent parameters. From this point of view, an appealing choice is the SO(10) group, where one generation of the SM fermion fields can be accommodated within one spinorial ${\bf 16}$ representation together with a right-handed neutrino. 

The SO(10) group has often been considered in the context of supersymmetry. In this context, SO(10) can be directly broken to the SM gauge group and such models have been studied for example in Refs.~\cite{Grimus:2007ye,Dutta:2005ni,Grimus:2006rk}. Without the assumption of supersymmetry, unification is still possible with the additional constraint that there exists an intermediate energy scale below $M_{\rm GUT}$. Thus, in the non-supersymmetric case, the symmetry must be broken at some high-energy scale (usually of $10^{16}$~GeV) into a gauge group of smaller rank, which successively breaks to the SM. The non-supersymmetric SO(10) models have been of little interest, since such were considered to be ruled out due to tachyonic masses in the scalar spectrum. This can be remedied by taking quantum effects into account, which again have made these models of interest \cite{Bertolini:2010ng}. Furthermore, there is no indication of supersymmetry in recent LHC data, which enhances the interest in the non-supersymmetric SO(10) models.

Any extension of the SM is viable only if the observables at the electroweak (EW) scale, $M_{\rm Z}$, can be accommodated within the model. Especially, this is true for the fermion masses and mixings. Many fits to fermion observables in the non-supersymmetric SO(10) model context have been performed assuming a simple SM running of the Yukawa couplings from the GUT scale $M_{\rm GUT}$ to the EW scale $M_{\rm Z}$ (or vice versa) \cite{Joshipura:2011nn,Altarelli:2013aqa,Dueck:2013gca} and have shown that the minimal choice of the Yukawa sector, which can accommodate all known low-energy experimental data, is the one with Higgses in the ${\bf 10}_{\rm H}$ and ${\bf 126}_{\rm H}$ representations. However, these did not take the effect of an intermediate energy scale into account. More recently, we have discussed the effects on such a renormalization group (RG) running given by a change in the gauge group at an intermediate energy scale $M_{\rm I}$ between $M_{\rm GUT}$ and $M_{\rm Z}$ \cite{Meloni:2014rga}, showing that this new energy threshold strongly affects the RG running of the Yukawa couplings from $M_{\rm GUT}$ to $M_{\rm Z}$, and thus, the final values of the EW observables.

The next-to-minimal choice is to enlarge the Yukawa sector of non-supersymmetric SO(10) models  adding another Higgs representation, ${\bf 120}_{\rm H}$. In the present work, we focus on models, where SO(10) is broken by the ${\bf 210}_{\rm H}$ \cite{Chang:1983fu,Chang:1984uy} to an intermediate gauge group, which is the Pati--Salam group ${\rm PS} = {\rm SU(4)}_{C} \otimes {\rm SU(2)}_{L} \otimes {\rm SU(2)}_{R}$ \cite{Pati:1974yy}. This model has recently been discussed in Ref.~\cite{Babu:2016bmy}. Given the fact that the ${\bf 120}_{\rm H}$ contains representations under the PS group such as $\left({\bf 1},{\bf 2},{\bf 2}\right)$ and $({\bf 15},{\bf 2},{\bf 2})$, it enters not only the relations defining the fermion masses but also affects the gauge coupling unification from $M_{\rm I}$ to $M_{\rm GUT}$ if we assume that the whole ${\bf 120}_{\rm H}$ multiplet has a mass around $M_{\rm I}$. In this model, neutrino mass is generated through the type-I seesaw mechanism, where the seesaw scale coincides with the intermediate scale. The relevant question that we want to address is whether or not such a model is able to describe all low-energy data on fermion masses and mixings, even in the presence of the intermediate gauge group, and to quantify how large the difference would be if a pure SM extrapolation of the SO(10) Yukawa couplings from $M_{\rm GUT}$ to $M_{\rm Z}$ is performed.

The paper is organized as follows. In Sec.~\ref{sec:RGEs}, we present the renormalization group equations (RGEs) and matching conditions for the gauge couplings, the Yukawa couplings, the Higgs self-couplings, and the effective neutrino mass matrix in the extended non-supersymmetric SO(10) model with an intermediate scale $M_{\rm I}$. Then, in Sec.~\ref{sec:numerics}, we discuss the numerical parameter-fitting procedure, which we use to perform the fit and the RG evolution from $M_{\rm GUT}$ to $M_{\rm Z}$. Next, in Sec.~\ref{sec:results}, we state our results for the RG running of the fermion masses and mixings, comparing the result to a SM-like model. In Sec.~\ref{sec:comparison}, we compare our results of the extended SO(10) model to the minimal SO(10) model. Finally, in Sec.~\ref{summary}, we summarize and conclude. In App.~\ref{sec:minimal}, we give the RGEs in the minimal SO(10) model, taking the opportunity to correct some errors in our and other previous works \cite{Fukuyama:2002vv,Meloni:2014rga}.

\section{Renormalization Group Equations of the Extended SO(10) Model}
\label{sec:RGEs}

In this section, the extended non-supersymmetric SO(10) model is presented including the RGEs and matching conditions of this model. Note that the minimal non-supersymmetric SO(10) model is discussed in App.~\ref{sec:minimal}. We evolve the RGEs to leading order in perturbation theory from the scale of a GUT $M_{\rm GUT}$ via an intermediate scale $M_{\rm I}$ down to the EW scale  $M_{\rm Z} = (91.1876 \pm 0.0021)$~GeV \cite{Agashe:2014kda}. The values of $M_{\rm GUT}$, $M_{\rm I}$, and $\alpha_{\rm GUT}^{-1}$ are determined by the running of the gauge couplings and the requirement that the gauge couplings should unify at $M_{\rm GUT}$. In the present case, we find 
\begin{equation}
M_{\rm GUT} = (1.0 \pm 0.5) \cdot 10^{16}~{\rm GeV}\,, \quad M_{\rm I} = (4.8 \pm 2.2) \cdot 10^{11}~{\rm GeV}\,, \quad\mbox{and}\quad \alpha_{\rm GUT}^{-1} = 28.6\,.
\label{eq:IC_GUT_E}
\end{equation}
Note that the values of $M_{\rm GUT}$ and $M_{\rm I}$ can be computed using Eqs.~(\ref{eq:2L})--(\ref{eq:4C}) and (\ref{eq:g1})--(\ref{eq:g3}) as well as the matching conditions~(\ref{eq:g1MZ})--(\ref{eq:g3MZ}), since the one-loop order RGEs for the gauge couplings only depend on the different gauge couplings \cite{Machacek:1983tz}.

\subsection{The Yukawa Lagrangian of the Extended SO(10) Model and Matching Conditions at $M_{\rm GUT}$}
\label{sec:matchingGaugeGUT}

Now, we present the Yukawa sector of the Lagrangian for the extended SO(10) model above and below $M_{\rm GUT}$ as well as the matching conditions for the Yukawa and gauge couplings at $M_{\rm GUT}$. We want to accommodate the ${\bf 120}_{\rm H}$ Higgs field. At the SO(10) level, i.e.~above $M_{\rm GUT}$, the Yukawa sector of the Lagrangian reads
\begin{equation}
-{\cal L}_Y^{\rm GUT} = {\bf 16}_{F} \left(h\, {\bf 10}_{\rm H} + f\, \overline{{\bf 126}}_{\rm H} + g\, {\bf 120}_{\rm H} \right){\bf 16}_F\,,
\end{equation}
where $h \equiv Y^{(10)}$, $f \equiv Y^{(126)}$, and $g \equiv Y^{(120)}$ are the Yukawa couplings. The couplings $f$ and $g$ can be represented by a symmetric and an antisymmetric matrix in flavor space, respectively, whereas $h$ can be represented by a real diagonal matrix. The Higgs representation ${\bf 10}_{\rm H}$ is real from the SO(10) perspective but its components could be chosen either real or complex. However, choosing the components of this representation real, i.e.~${\bf 10}_{\rm H}={\bf 10}_{\rm H}^*$, would imply that $k_u=k_d$. This issue can be solved by complexifying this representation and in addition introducing a Peccei--Quinn symmetry, ${\rm U(1)}_{\rm PQ}$, which then forbids Yukawa couplings with a ${\bf 10}_{\rm H}^*$ \cite{Babu:1992ia}. The Peccei--Quinn symmetry is defined as
\[
{\bf 16}_{F} \rightarrow e^{{\rm i}\alpha}{\bf 16}_{F}\,, \quad {\bf 10}_{\rm H} \rightarrow e^{-2{\rm i} \alpha}{\bf 10}_{\rm H}\,, \quad\overline{{\bf 126}}_{\rm H} \rightarrow e^{-2{\rm i} \alpha}\overline{{\bf 126}}_{\rm H}\,, \quad\overline{{\bf 120}}_{\rm H} \rightarrow e^{-2{\rm i} \alpha}\overline{{\bf 120}}_{\rm H}\,,
\]
where $\alpha$ is a phase and represents the ${\rm U(1)}_{\rm PQ}$ charge. Under the PS group, the ${\bf 120}_{\rm H}$ decomposes as
\begin{equation}
{\bf 120}_{\rm H} = ({\bf 10}+\overline{{\bf 10}},{\bf 1},{\bf 1}) \oplus ({\bf 6},{\bf 3},{\bf 1})\oplus ({\bf 6},{\bf 1},{\bf 3}) 
\oplus ({\bf 15},{\bf 2},{\bf 2})\oplus ({\bf 1},{\bf 2},{\bf 2})\,,
\end{equation}
which means that we have four bidoublets that take part in generating the fermion masses. For the ${\bf 126}_{\rm H}$, we assume that they are close to $M_{\rm I}$. We introduce the following simplifying abbreviations for the fields
\begin{equation}
\Phi_{120} \equiv \left({\bf 1},{\bf 2},{\bf 2}\right) \,, \qquad \Sigma_{120} \equiv ({\bf 15},{\bf 2},{\bf 2}) \,.
\end{equation}
At the PS level, i.e.~below $M_{\rm GUT}$ and above $M_{\rm I}$, the Yukawa sector of the Lagrangian for the extended SO(10) model then reads
\begin{align}
- {\mathcal L}_Y^{10 + \overline{126}+120} &= Y_F^{(10)} \bar F_L \Phi_{10} F_R + Y_F^{(126)} \bar F_L \Sigma_{126} F_R + 
Y_R^{(126)} F_R^{\rm T} C F_R \overline{\Delta_R} \nonumber \\
&+Y_{F,1}^{(120)} \bar F_L \Phi_{120} F_R + Y_{F,2}^{(120)} \bar F_L \Sigma_{120} F_R \,,
\label{eq:y3}
\end{align}
where $Y_F^{(10)}$ and $Y_F^{(126)}$ are the Yukawa couplings of the SU(4) singlet and the SU(4) $15$-plet of the 
${\bf 10}_{\rm H}$ and ${\bf 126}_{\rm H}$, respectively, $Y_{F,1}^{(120)}$ and  $Y_{F,2}^{(120)}$ are the analogous for the 
${\bf 120}_{\rm H}$, $Y_R^{(126)}$ is the right-handed Majorana neutrino coupling, 
and $F_L = ({\bf 4},{\bf 2},{\bf 1})$, $F_R = ({\bf 4},{\bf 1},{\bf 2})$, $\Phi_{10} = ({\bf 1},{\bf 2},{\bf 2})$, $\Sigma_{126} = ({\bf 15},{\bf 2},{\bf 2})$, and $\overline{\Delta_R} = (\overline{{\bf 10}},{\bf 1},{\bf 3})$. Therefore, at $M_{\rm GUT}$, we need to match the two different sets of Yukawa couplings above and below $M_{\rm GUT}$. These matching conditions are \cite{Aulakh:2002zr,Fukuyama:2002vv,Meloni:2014rga}
\begin{align}
\frac{1}{\sqrt{2}}Y^{(10)}_F (M_{\rm GUT}) &\equiv h\,, \label{eq:bc1} \\
\frac{1}{4\sqrt{2}}Y^{(126)}_F(M_{\rm GUT}) =\frac{1}{4}Y^{(126)}_R (M_{\rm GUT})
&\equiv f\,, \label{bc2}\\
\frac{1}{\sqrt{2}}Y^{(120)}_{F,1} (M_{\rm GUT}) = - \frac{1}{2\sqrt{2}}Y^{(120)}_{F,2} (M_{\rm GUT}) & \equiv g\,.
\end{align}

In addition, at $M_{\rm GUT}$, the gauge coupling constants unify, and therefore, the matching conditions are simply
\begin{equation}
g_{2L}(M_{\rm GUT}) \equiv g_{2R}(M_{\rm GUT}) \equiv g_{4C}(M_{\rm GUT}) \,,
\label{eq:matchingGaugeGUT}
\end{equation}
where $g_{2L}$, $g_{2R}$, and $g_{4C}$ are the ${\rm SU(2)}_{L}$, ${\rm SU(2)}_{R}$, and ${\rm SU(4)}_{C}$ gauge coupling constants, respectively. Note that the value of the gauge coupling at $M_{\rm GUT}$ is given by
\begin{equation}
\alpha_{\rm GUT} = 0.035 \,,
\end{equation}
which follows directly from Eq.~\eqref{eq:IC_GUT_E}.

\subsection{RGEs from $M_{\rm GUT}$ to $M_{\rm I}$}

In this subsection, we give the RGEs for the gauge and Yukawa couplings between $M_{\rm GUT}$ and $M_{\rm I}$, see Secs.~\ref{sec:RGEsGC} and \ref{sec:Yukawa_above}, respectively.

\subsubsection{RGEs for the Gauge Couplings}\label{sec:RGEgaugecoupling_above}
\label{sec:RGEsGC}

Between $M_{\rm GUT}$ and $M_{\rm I}$, the RGEs for the gauge coupling constants $g_{2L}$, $g_{2R}$, and $g_{4C}$ read
\begin{align}
16 \pi^2 \frac{{\rm d} g_{2L}}{{\rm d} t} & =  \frac{22}{3} g_{2L}^3 \,, \label{eq:2L} \\
16 \pi^2 \frac{{\rm d} g_{2R}}{{\rm d} t} & = 14 g_{2R}^3 \,, \label{eq:2R} \\
16 \pi^2 \frac{{\rm d} g_{4C}}{{\rm d} t} & = 3 g_{4C}^ 3 \,, \label{eq:4C}
\end{align}
where $t \equiv \ln \mu$ ($\mu$ being the energy scale), which have the evolutions between two energy scales $M_1$ and $M_2$ given by the standard formula \cite{Jones:1981we,Koh:1983ir}
\begin{equation}
g_i^{-2}(M_2) = g_i^{-2}(M_1) - \frac{a_i}{8 \pi^2} \log \left(\frac{M_2}{M_1}\right) \,,
\end{equation}
where the coefficients $a_i$ can be obtained from e.g.~Ref.~\cite{Koh:1983ir} and are listed in Eqs.~(\ref{eq:2L})--(\ref{eq:4C}). 

\subsubsection{RGEs for the Yukawa Couplings}
\label{sec:Yukawa_above}

Furthermore, between $M_{\rm GUT}$ and $M_{\rm I}$, the RGEs for the Yukawa couplings $Y_F^{(10)}$, $Y_F^{(126)}$, $Y_R^{(126)}$, $Y_{F,1}^{(120)}$, and $Y_{F,2}^{(120)}$ are found to be
{\allowdisplaybreaks  \begin{align} 
16 \pi^2 \frac{{\rm d} Y_F^{(10)}}{{\rm d} t} & =  
\Big\{{Y_F^{(10)}}  {Y_F^{(10)}}^{\dagger} + {Y_{F,1}^{(120)}}  {Y_{F,1}^{(120)}}^{\dagger} + \frac{15}{4} \Big({{Y_F^{(126)}}  {Y_F^{(126)}}^{\dagger} + {Y_{F,2}^{(120)}}  {Y_{F,2}^{(120)}}^{\dagger}\Big) \Big\} {Y_F^{(10)}}} \nonumber\\
&+ {Y_F^{(10)}} \Big\{ {Y_F^{(10)}}^{\dagger}  {Y_F^{(10)}} + {Y_{F,1}^{(120)}}^{\dagger}  {Y_{F,1}^{(120)}} \nonumber\\
&+ \frac{15}{4} \Big({Y_F^{(126)}}^{\dagger}  {Y_F^{(126)}} + {Y_{F,2}^{(120)}}^{\dagger}  {Y_{F,2}^{(120)}} + 2 {Y_R^{(126)}}^*  {Y_R^{(126)}}\Big)\Big\}\nonumber\\
& + 4 \, \mbox{tr}\Big({{Y_F^{(10)}}  {Y_F^{(10)}}^{\dagger}}\Big) {Y_F^{(10)}} + 4 \, \mbox{tr}\Big({{Y_F^{(10)}}  {Y_{F,1}^{(120)}}^{\dagger}}\Big) {Y_{F,1}^{(120)}} - \frac{9}{4} \Big(g_{2L}^{2} + g_{2R}^{2} + 5 g_{4C}^{2}\Big) {Y_F^{(10)}} \,,\label{eq:YF10}\\ 
16 \pi^2 \frac{{\rm d} Y_F^{(126)}}{{\rm d} t} & =  
\Big\{{{Y_F^{(10)}}}  {Y_F^{(10)}}^{\dagger} + {Y_{F,1}^{(120)}}  {Y_{F,1}^{(120)}}^{\dagger} + \frac{15}{4} \Big({{Y_F^{(126)}}  {Y_F^{(126)}}^{\dagger} + {Y_{F,2}^{(120)}}  {Y_{F,2}^{(120)}}^{\dagger}\Big)\Big\} {Y_F^{(126)}}} \nonumber\\
&+ {Y_F^{(126)}} \Big\{ {Y_F^{(10)}}^{\dagger}  
{Y_F^{(10)}} + {Y_{F,1}^{(120)}}^{\dagger}  {Y_{F,1}^{(120)}} \nonumber\\
&+ \frac{15}{4} \Big({Y_F^{(126)}}^{\dagger}  {Y_F^{(126)}} + {Y_{F,2}^{(120)}}^{\dagger}  {Y_{F,2}^{(120)}} + 2 {Y_R^{(126)}}^*  {Y_R^{(126)}}\Big)\Big\}\nonumber\\
& + \mbox{tr}\Big({{Y_F^{(126)}} 
{Y_F^{(126)}}^{\dagger}}\Big) {Y_F^{(126)}} + \mbox{tr}\Big({{Y_F^{(126)}}  {Y_{F,2}^{(120)}}^{\dagger}}\Big) {Y_{F,2}^{(120)}} - \frac{9}{4} \Big(g_{2L}^{2} + g_{2R}^{2} + 5 g_{4C}^{2}\Big) {Y_F^{(126)}} \,,\label{eq:YF126}\\ 
16 \pi^2 \frac{{\rm d} Y_R^{(126)}}{{\rm d} t} & =  
\Big\{{Y_F^{(10)}}^{T}  {Y_F^{(10)}}^* + {Y_{F,1}^{(120)}}^{T}  {Y_{F,1}^{(120)}}^* \nonumber\\
&+ \frac{15}{4} \Big({{Y_F^{(126)}}^{T}  {Y_F^{(126)}}^* + {Y_{F,2}^{(120)}}^{T}  {Y_{F,2}^{(120)}}^* + 2 {{Y_R^{(126)}}}  {Y_R^{(126)}}^*\Big)\Big\} {Y_R^{(126)}}}
\nonumber\\
& + {{Y_R^{(126)}}}  \Big\{{Y_F^{(10)}}^{\dagger}  {Y_F^{(10)}} + {Y_{F,1}^{(120)}}^{\dagger}  {Y_{F,1}^{(120)}} \nonumber\\
&+ \frac{15}{4} \Big({Y_F^{(126)}}^{\dagger}  {Y_F^{(126)}} + {Y_{F,2}^{(120)}}^{\dagger}  {Y_{F,2}^{(120)}} + 2 {Y_R^{(126)}}^* {Y_R^{(126)}}\Big)\Big\}\nonumber\\
& + 2 \, \mbox{tr}\Big({{Y_R^{(126)}}  {Y_R^{(126)}}^*}\Big) {Y_R^{(126)}} -
\frac{9}{4} \Big(2 g_{2R}^{2}  + 5 g_{4C}^{2}\Big) {Y_R^{(126)}} \,, \label{eq:YR126}\\ 
16 \pi^2 \frac{{\rm d} Y_{F,1}^{(120)}}{{\rm d} t} & = 
\Big\{{Y_{F}^{(10)}}  {Y_{F}^{(10)}}^{\dagger} + {Y_{F,1}^{(120)}}  {Y_{F,1}^{(120)}}^{\dagger} + \frac{15}{4} \Big( {Y_{F}^{(126)}}  {Y_{F}^{(126)}}^{\dagger} + 
{Y_{F,2}^{(120)}}  {Y_{F,2}^{(120)}}^{\dagger}\Big)\Big\} {Y_{F,1}^{(120)}} \nonumber\\ & + {Y_{F,1}^{(120)}} \Big\{ {Y_{F}^{(10)}}^{\dagger} 
{Y_{F}^{(10)}} + {Y_{F,1}^{(120)}}^{\dagger}  {Y_{F,1}^{(120)}} \nonumber\\
&+ \frac{15}{4} \Big({Y_{F}^{(126)}}^{\dagger}  {Y_{F}^{(126)}} + {Y_{F,2}^{(120)}}^{\dagger}  
{Y_{F,2}^{(120)}} + 2 {Y_R^{(126)}}^*  {Y_R^{(126)}}\Big)\Big\}\nonumber\\ & + 4 \, \mbox{tr}\Big({{Y_{F,1}^{(120)}}  {Y_{F,1}^{(120)}}^{\dagger}}\Big) 
{Y_{F,1}^{(120)}} + 4 \, \mbox{tr}\Big({{Y_{F,1}^{(120)}}  {Y_{F}^{(10)}}^{\dagger}}\Big) {Y_{F}^{(10)}} - \frac{9}{4} \Big(g_{2L}^{2} + g_{2R}^{2} + 5 g_{4C}^{2}\Big)
{Y_{F,1}^{(120)}} \,, \label{eq:YF1}\\ 
16 \pi^2 \frac{{\rm d} Y_{F,2}^{(120)}}{{\rm d} t} & = 
\Big\{{Y_{F}^{(10)}}  {Y_{F}^{(10)}}^{\dagger} + {Y_{F,1}^{(120)}}  {Y_{F,1}^{(120)}}^{\dagger} + \frac{15}{4} \Big( {Y_{F}^{(126)}}  {Y_{F}^{(126)}}^{\dagger} + 
{Y_{F,2}^{(120)}}  {Y_{F,2}^{(120)}}^{\dagger}\Big)\Big\} {Y_{F,2}^{(120)}}\nonumber\\ & + {Y_{F,2}^{(120)}} \Big\{ {Y_{F}^{(10)}}^{\dagger}  {Y_{F}^{(10)}} + 
{Y_{F,1}^{(120)}}^{\dagger}  {Y_{F,1}^{(120)}} \nonumber\\
&+ \frac{15}{4} \Big({Y_{F}^{(126)}}^{\dagger}  {Y_{F}^{(126)}} + {Y_{F,2}^{(120)}}^{\dagger}  {Y_{F,2}^{(120)}} +
2 {Y_R^{(126)}}^*  {Y_R^{(126)}}\Big)\Big\}\nonumber\\ & + \mbox{tr}\Big({{Y_{F,2}^{(120)}}  {Y_{F,2}^{(120)}}^{\dagger}}\Big) {Y_{F,2}^{(120)}} + 
\mbox{tr}\Big({{Y_{F,2}^{(120)}}  {Y_{F}^{(126)}}^{\dagger}}\Big) {Y_{F}^{(126)}} - \frac{9}{4} \Big(g_{2L}^{2} + g_{2R}^{2} + 5 g_{4C}^{2}\Big) {Y_{F,2}^{(120)}}
\label{eq:YF2}\,.
\end{align}}
Note that Eqs.~(\ref{eq:2L})--(\ref{eq:4C}) and (\ref{eq:YF10})--(\ref{eq:YF2}) have been computed using the software {\sc SARAH 4} \cite{Staub:2013tta}. We briefly discuss the different parts of the RGEs for the Yukawa couplings. The three first lines of Eqs.~\eqref{eq:YF10}--\eqref{eq:YF2} (actually the four first lines of Eq.~\eqref{eq:YR126}) stem from self-energies of the fermion fields $F_L$ and $F_R$. Note that the self-energies of $F_L$ and $F_R$ differ by a term containing $Y_R^{(126)}$, since the field $\overline{\Delta_R}$ does not couple to $F_L$. The two first terms in the last lines of Eqs.~\eqref{eq:YF10}--\eqref{eq:YF2} come from fermion-loop contributions to the self-energies of the scalars. Finally, the last terms of Eqs.~\eqref{eq:YF10}--\eqref{eq:YF2} are contributions from the gauge couplings to the RGEs. Note that there is a difference in the gauge coupling term of Eq.~\eqref{eq:YR126}, which again stems from the fact $\overline{\Delta_R}$ does not couple to $F_L$. The corrections to the Yukawa coupling vertices do not give any contributions to the RGEs.

In the extended SO(10) model, there are two $({\bf 1},{\bf 2},{\bf 2})$ representations in the ${\bf 10}_{\rm H}$ and ${\bf 120}_{\rm H}$, respectively, i.e.~$\Phi_{10}$ and $\Phi_{120}$, which couple identically to the other representations. The same applies to the two $({\bf 15},{\bf 2},{\bf 2})$ representations in the ${\bf 126}_{\rm H}$ and ${\bf 120}_{\rm H}$, respectively, i.e.~$\Sigma_{126}$ and $\Sigma_{120}$. As a consequence, we can observe that Eq.~(\ref{eq:YF1}) can be obtained from Eq.~(\ref{eq:YF10}) by making the replacement $Y_F^{(10)} \to Y_{F,1}^{(120)}$. Similarly, Eq.~(\ref{eq:YF2}) can be obtained from Eq.~(\ref{eq:YF126}) by making the replacement $Y_F^{(126)} \to Y_{F,2}^{(120)}$. Naturally, Eq.~\eqref{eq:YF1} cannot be obtained directly from Eq.~(\ref{eq:YF126}). However, there are only minor differences in these two equations. To be precise, the only difference is the factor of 4 in front of the traces, which is due to their respective group theoretical structure under ${\rm SU(4)}_{C}$, where the representation is either trivial or adjoint. Equation~\eqref{eq:YR126} differs from the other RGEs, since $Y_R^{(126)}$ couples to $F_R$ only.

\subsection{Matching Conditions at $M_{\rm I}$}\label{sec:matching}

In this subsection, we display the matching conditions for the gauge and Yukawa couplings as well as the quartic scalar Higgs self-coupling and the effective neutrino mass matrix at $M_{\rm I}$, see Secs.~\ref{sec:matchingGauge}, \ref{sec:matchYC}, \ref{sec:matchQSHSC}, and \ref{sec:matchENMM}, respectively.

\subsubsection{Matching Conditions for the Gauge Couplings}
\label{sec:matchingGauge}

At $M_{\rm I}$, we impose the following matching conditions for the gauge coupling constants \cite{Hall:1980kf,Deshpande:1992au,Fukuyama:2002vv,Meloni:2014rga}
\begin{align}
g_1^{-2}(M_{\rm I}) &\equiv \frac{3}{5} g_{2R}^{-2}(M_{\rm I}) + \frac{2}{5} g_{4C}^{-2}(M_{\rm I}) \,, \\
g_2(M_{\rm I}) &\equiv g_{2L}(M_{\rm I}) \,, \\
g_3(M_{\rm I}) &\equiv g_{4C}(M_{\rm I}) \,,
\end{align}
where $g_1$, $g_2$, and $g_3$ are the gauge coupling constants of U(1), ${\rm SU(2)}_L$, and ${\rm SU(3)}_C$, respectively. 

\subsubsection{Matching Conditions for the Yukawa Couplings}
\label{sec:matchYC}

At $M_{\rm GUT}$, the fermion mass matrices are defined as
\begin{align}
M_u &= h k_u + f v_u + g\, (t_u + z_u) \,, \label{eq:Mu}\\
M_d &= h k_d + f v_d + g \,z_d \,, \\
M_{\nu_D} &= h k_u - 3 f v_u + g\, (-3 t_u +z_u) \,, \\
M_e &= h k_d - 3 f v_d + g\, t_d \,, \label{eq:Me}
\end{align}
where the vacuum expectation values (VEVs) are defined as
\begin{equation}\label{eq:vevs}
k_{u,d} = \langle \Phi_{10}\rangle_{u,d} \,, \qquad v_{u,d} = \langle \Sigma_{126}\rangle_{u,d} \,, \qquad z_{u,d} = \langle \Phi_{120}\rangle_{u,d} \,, \qquad t_{u,d} = \langle \Sigma_{120}\rangle_{u,d} \,.
\end{equation}
Adopting a rescaling of the VEVs, Eqs.~(\ref{eq:Mu})--(\ref{eq:Me}) can be recast in the following way
\begin{align}
M_u &= \frac{r_v}{\sqrt{2}} \left(k_d Y^{(10)}_F + \frac{v_d s}{4} Y^{(126)}_F\right) + \frac{1}{\sqrt{2}}
\left(z_u Y^{(120)}_{F,1} - \frac{t_u}{2} Y^{(120)}_{F,2}\right) \,, \\
M_d &= \frac{k_d}{\sqrt{2}} Y^{(10)}_F + \frac{v_d}{4\sqrt{2}} Y^{(126)}_F + \frac{z_d}{\sqrt{2}}
Y^{(120)}_{F,1} \,, \\
M_{\nu_D} &= \frac{r_v}{\sqrt{2}} \left(k_d Y^{(10)}_F - \frac{3 v_d s}{4} Y^{(126)}_F\right) + \frac{1}{\sqrt{2}}
\left(z_u Y^{(120)}_{F,1} +\frac{3 t_u}{2} Y^{(120)}_{F,2}\right) \,, \\
M_e &= \frac{k_d}{\sqrt{2}} Y^{(10)}_F - \frac{3 v_d}{4\sqrt{2}} Y^{(126)}_F - \frac{t_d}{2\sqrt{2}} Y^{(120)}_{F,2} \,.
\end{align}
Furthermore, $r_v \equiv k_u / k_d$ and $s \equiv v_u / (r_v v_d)$.
Now, we need the matching conditions at $M_{\rm I}$. Since at the EW level, i.e.~below $M_{\rm I}$ and above $M_{\rm Z}$, the Yukawa sector of the Lagrangian for a two-Higgs-doublet model is given by
\begin{equation}
- {\mathcal L}_{\rm Y}^{\rm 2HDM} = Y_u\bar{q}_L \phi_2 u_R + Y_d\bar{q}_L \phi_1 d_R +Y_e\bar{\ell}_L \phi_1 e_R \,,
\label{eq:y}
\end{equation}
where $Y_u$, $Y_d$, and $Y_e$ are three Yukawa couplings, we have
\begin{equation}
M_u = Y_u \frac{k_u}{\sqrt{2}} \,, \qquad M_e =Y_e \frac{k_d}{\sqrt{2}} \,, \qquad M_d =Y_d \frac{k_d}{\sqrt{2}} \,.
\end{equation}
Thus, at $M_{\rm I}$, the matching produces the following relations
\begin{align}
Y_u (M_{\rm I}) &\equiv Y_F^{(10)} (M_{\rm I}) + \frac{s v_d}{4 k_d} Y^{(126)}_F (M_{\rm I}) + \frac{z_u}{k_d r_v} Y^{(120)}_{F,1} (M_{\rm I}) - \frac{t_u}{2 k_d r_v} Y^{(120)}_{F,2} (M_{\rm I})\,, \label{eq:YuMI}\\
Y_d (M_{\rm I}) &\equiv Y_F^{(10)} (M_{\rm I}) + \frac{v_d}{4 k_d} Y^{(126)}_F (M_{\rm I}) + \frac{z_d}{k_d} Y^{(120)}_{F,1} (M_{\rm I})\,, \label{eq:YdMI}\\
Y_e (M_{\rm I}) &\equiv Y_F^{(10)} (M_{\rm I}) - \frac{3 v_d}{4 k_d} Y^{(126)}_F (M_{\rm I}) - \frac{t_d}{2 k_d} Y^{(120)}_{F,2} (M_{\rm I})\,. \label{eq:YeMI}
\end{align}
Note that the corresponding matching conditions for the minimal SO(10) model are obtained from Eqs.~\eqref{eq:YuMI}--\eqref{eq:YeMI} by setting $z_{u,d}=t_{u,d}=0$, cf.~Eqs.~\eqref{eq:matchmi1}--\eqref{eq:matchmi3}.

\subsubsection{Matching Conditions for the Quartic Scalar Higgs Self-Couplings}
\label{sec:matchQSHSC}

Using the two-Higgs-doublet model, we assume for simplicity the following Higgs potential below $M_{\rm I}$
\begin{align}
V = \lambda_1\left(\phi_1^\dagger\phi_1\right)\left(\phi_1^\dagger\phi_1\right)+\lambda_2\left(\phi_2^\dagger\phi_2\right)\left(\phi_2^\dagger\phi_2\right) \,,
\label{eq:Vextended}
\end{align}
where $\lambda_1$ and $\lambda_2$ are the two quartic scalar Higgs self-couplings of the model.
Note that we use the so-called GUT normalization, which means that $g_1 = g' \sqrt{5/3}$, where $g_1$ is the U(1) gauge coupling constant with a normalization based on SU(5) and $g'$ is the standard EW coupling constant. At $M_{\rm I}$, for the matching conditions of $\lambda_1$ and $\lambda_2$, we will assume 
\begin{align}
\lambda_1(M_{\rm I}) \equiv \lambda_2(M_{\rm I}) \equiv \rm{const.}
\end{align}

\subsubsection{Matching Condition for the Effective Neutrino Mass Matrix}
\label{sec:matchENMM}

The neutrino masses are generated through a type-I seesaw mechanism, for simplicity we assume that the seesaw scale coincides with $M_{\rm I}$. We have to make a matching for the effective neutrino mass matrix at $M_{\rm I}$. Below $M_{\rm I}$, the effective neutrino mass matrix is given by
\begin{equation}
m_\nu = \frac{\kappa k_u^2}{2} \,.
\label{eq:mkappa}
\end{equation}
At $M_{\rm I}$, this expression must match the form of $m_\nu$ valid from $M_{\rm GUT}$ to $M_{\rm I}$, which can be conveniently expressed as
\begin{align}
m_\nu &= \left(4 k_d r_v Y_F^{(10)T}-3 r_v s
   v_d Y_F^{(126)T}+6 t_u Y_{F,2}^{(120)T}+4 Y_{F,1}^{(120)T} z_u\right)\nonumber \\
&\times \left(32 M_R\right)^{-1}
\left(4 k_d r_v Y_F^{(10)}-3 r_v s v_d Y_F^{(126)}+6 t_u Y_{F,2}^{(120)}+4 Y_{F,1}^{(120)} z_u\right) \,,
\end{align}
where $M_R$ is a RG running quantity defined as
\begin{equation}
M_R \equiv \frac{1}{4}\left\langle \overline{\Delta_R} \right\rangle Y^{(126)}_R\,.
\label{triplet}
\end{equation}
Therefore, at $M_{\rm I}$, we have the following matching condition
\begin{align}
\kappa(M_{\rm I}) &\equiv \left\{4 k_d r_v Y_F^{(10)T}(M_{\rm I}) - 3 r_v s
   v_d Y_F^{(126)T}(M_{\rm I}) + 6 t_u Y_{F,2}^{(120)T}(M_{\rm I}) + 4 z_u Y_{F,1}^{(120)T}(M_{\rm I}) \right\}\nonumber \\
&\times \left\{16\,k_u^2\, M_R(M_{\rm I})\right\}^{-1}
\left\{4 k_d r_v Y_F^{(10)}(M_{\rm I}) - 3 r_v s v_d Y_F^{(126)}(M_{\rm I}) + 6 t_u Y_{F,2}^{(120)}(M_{\rm I}) + 4 z_u Y_{F,1}^{(120)}(M_{\rm I}) \right\} \,.
\end{align}

\subsection{RGEs from $M_{\rm I}$ to $M_{\rm Z}$}

In this subsection, we give the RGEs for the gauge and Yukawa couplings, the quartic scalar Higgs self-coupling, and the effective neutrino mass matrix between $M_{\rm I}$ and $M_{\rm Z}$, see Secs.~\ref{sec:RGEgaugecoupling}, \ref{sec:Yukawa_below}, \ref{sec:RGEsQSHSC}, and \ref{sec:RGEsENMM}, respectively.

\subsubsection{RGEs for the Gauge Couplings}
\label{sec:RGEgaugecoupling}

Below $M_{\rm I}$, we assume that there are two Higgs doublets, in which case the RGEs for the gauge couplings $g_1$, $g_2$, and $g_3$ read \cite{Jones:1981we}
\begin{align}
16 \pi^2 \frac{{\rm d} g_1}{{\rm d} t} & = \frac{21}{5} g_1^3 \,, \label{eq:g1} \\
16 \pi^2 \frac{{\rm d} g_2}{{\rm d} t} & = - 3 g_2^3 \,, \label{eq:g2} \\
16 \pi^2 \frac{{\rm d} g_3}{{\rm d} t} & = - 7 g_3^3 \,. \label{eq:g3}
\end{align}

\subsubsection{RGEs for the Yukawa Couplings}
\label{sec:Yukawa_below}

The Higgs doublets couples to the lepton and quark fields according to Eq.~\eqref{eq:y}. Thus, below $M_{\rm I}$, we have three Yukawa couplings $Y_u$, $Y_d$, and $Y_e$, and for the RGEs of these Yukawa couplings, we obtain
{\allowdisplaybreaks  \begin{align} 
16 \pi^2 \frac{{\rm d} Y_u}{{\rm d} t} & =  
3 \mbox{tr}\Big({Y_u  Y_{u}^{\dagger}}\Big) Y_u - \Big(8 g_{3}^{2} + \frac{9}{4} g_{2}^{2} + \frac{17}{20} g_{1}^{2} \Big) Y_u + {Y_u  Y_{u}^{\dagger}  Y_u} + \frac{1}{2} Y_u \Big( {Y_{u}^{\dagger}  Y_u}  + {Y_{d}^{\dagger}  Y_d}\Big) \,, \label{eq:Yu}\\ 
16 \pi^2 \frac{{\rm d} Y_d}{{\rm d} t} & =  
\Big\{ 3 \mbox{tr}\Big({Y_d  Y_{d}^{\dagger}}\Big) + \mbox{tr}\Big({Y_e  Y_{e}^{\dagger}}\Big) \Big\}Y_d - \Big( 8 g_{3}^{2} + \frac{9}{4} g_{2}^{2} + \frac{1}{4} g_{1}^{2} \Big) Y_d + {Y_d  Y_{d}^{\dagger}  Y_d} + \frac{1}{2} Y_d \Big( {Y_{u}^{\dagger}  Y_u}  + {Y_{d}^{\dagger}  Y_d} \Big) \,,\label{eq:Yd}\\ 
16 \pi^2 \frac{{\rm d} Y_e}{{\rm d} t} & =  
\Big\{ 3 \mbox{tr}\Big({Y_d  Y_{d}^{\dagger}}\Big) + \mbox{tr}\Big({Y_e  Y_{e}^{\dagger}}\Big) \Big\}Y_e - \Big( \frac{9}{4} g_{2}^{2} + \frac{9}{4} g_{1}^{2}\Big) Y_e + \frac{3}{2} {Y_e  Y_{e}^{\dagger}  Y_e} \,. \label{eq:Ye}
\end{align}}
Note that Eqs.~(\ref{eq:Yu})--(\ref{eq:Ye}) have been computed using the software {\sc SARAH 4} \cite{Staub:2013tta}. Similar RGEs to Eqs.~(\ref{eq:Yu})--(\ref{eq:Ye}) have been presented previously in the literature \cite{Branco:2011iw}.

\subsubsection{RGEs for the Quartic Scalar Higgs Self-Couplings}
\label{sec:RGEsQSHSC}

Then, below $M_{\rm I}$, the RGEs for $\lambda_1$ and $\lambda_2$ described in Eq.~\eqref{eq:Vextended} are given by \cite{Grimus:2004yh,Branco:2011iw}
{\allowdisplaybreaks  \begin{align} 
16 \pi^2 \frac{{\rm d} \lambda_{1}}{{\rm d} t} & = 24\lambda_1^2- \Big(9 g_2^2 + \frac{9}{5} g_1^2\Big) \lambda_1 + \frac{9}{8} g_2^4 + \frac{9}{20} g_2^2 g_1^2 + \frac{27}{200} g_1^4 \nonumber\\ & + 12 {\rm tr}(Y_uY_u^\dagger) \lambda_1 - 6{\rm tr}(Y_u Y_u^\dagger Y_u Y_u^\dagger) \,, \label{eq:lambda1}
\\ 
16 \pi^2 \frac{{\rm d} \lambda_{2}}{{\rm d} t} & = 24\lambda_2^2- \Big(9 g_2^2 + \frac{9}{5} g_1^2\Big) \lambda_2 + \frac{9}{8} g_2^4 + \frac{9}{20} g_2^2 g_1^2 + \frac{27}{200} g_1^4 \nonumber\\ & + 4 {\rm tr}(Y_e Y_e^\dagger+3Y_dY_d^\dagger) \lambda_2 - 2 {\rm tr}(Y_e Y_e^\dagger Y_e Y_e^\dagger) - 6{\rm tr}(Y_d Y_d^\dagger Y_d Y_d^\dagger) \,. \label{eq:lambda2}
\end{align}}
Note that we have also checked Eqs.~(\ref{eq:lambda1}) and (\ref{eq:lambda2}) by recomputing them with the software {\sc SARAH 4} \cite{Staub:2013tta}. The first line of Eq.~\eqref{eq:lambda2} can be obtained from the first line of Eq.~\eqref{eq:lambda1} by making the replacement $\lambda_1 \to \lambda_2$, and vice versa. The difference between Eqs.~\eqref{eq:lambda1} and \eqref{eq:lambda2} consists of the terms containing traces of the Yukawa couplings. The Yukawa couplings are given in Eq.~\eqref{eq:y} and the ones in the traces naturally depend on which Higgs doublet the up, down, and lepton doublet couple to. General formulas for the RGEs of quartic scalar Higgs self-couplings have been derived earlier, see for example Refs.~\cite{Cheng:1973nv,Komatsu:1981xh,Machacek:1984zw,Fukuyama:2002vv,Grimus:2004yh,Branco:2011iw,Meloni:2014rga}. Some of these derivations are more correct than others.

\subsubsection{RGE for the Effective Neutrino Mass Matrix}
\label{sec:RGEsENMM}

Finally, below $M_{\rm I}$, in the case of the two-Higgs-doublet model, the RGE for the effective neutrino mass matrix $\kappa$ is given by \cite{Antusch:2001vn,Grimus:2004yh}
\begin{equation} 
16 \pi^2 \frac{{\rm d} \kappa}{{\rm d} t} = - 3 g_2^2 \kappa + 4 \lambda_2 \kappa + 6 {\rm tr} \Big(Y_d Y_d^\dagger\Big) \kappa + 2 {\rm tr} \Big(Y_e Y_e^\dagger\Big) \kappa - \frac{3}{2} \Big( \kappa Y_e^\dagger Y_e  + Y_e^T Y_e^* \kappa \Big) \,.
\end{equation}

\subsection{Matching Conditions at $M_{\rm Z}$}
\label{sec:matchMZ}

In this subsection, we show the matching conditions for the gauge couplings at $M_{\rm Z}$. At $M_{\rm Z}$, the experimental values of the gauge coupling constants [$\alpha_k = g_k^2/(4\pi)$] read \cite{Agashe:2014kda}
\begin{align}
g_1(M_{\rm Z}) &\equiv 0.463 \pm 0.001 \quad \mbox{or equivalently} \quad \alpha_1(M_{\rm Z}) = 0.0170 \pm 0.0001 \,, \label{eq:g1MZ} \\
g_2(M_{\rm Z}) &\equiv 0.654 \pm 0.001 \quad \mbox{or equivalently} \quad \alpha_2(M_{\rm Z}) = 0.0340 \pm 0.0001 \,, \label{eq:g2MZ} \\
g_3(M_{\rm Z}) &\equiv 1.220 \pm 0.003 \quad \mbox{or equivalently} \quad \alpha_3(M_{\rm Z}) = 0.1185 \pm 0.0006 \,, \label{eq:g3MZ}
\end{align}
which we impose as the matching conditions at $M_{\rm Z}$.

\section{Numerical Parameter-Fitting Procedure}\label{sec:numerics}

We perform a full numerical scan of the parameter space for the extended non-supersymmetric SO(10) model with an intermediate scale $M_{\rm I}$ such that $M_{\rm Z} \leq M_{\rm I} \leq M_{\rm GUT}$. In this model, there are in total 33 free parameters. The three Yukawa coupling matrices $h$, $f$, and $g$ consist of 21 parameters, which are three real parameters in the real diagonal matrix $h$, six complex parameters in the symmetric matrix $f$, and three complex parameters in the antisymmetric matrix $g$, respectively. In addition, there are eleven parameters related to the VEVs, i.e.~$r_v=k_u/k_d$, $t_d$, and $z_d$ (three real parameters) and $v_u$, $v_d$, $t_u$, and $z_u$ (four complex parameters). Finally, there is one parameter related to the two Higgs self-coupling constants $\lambda_1$ and $\lambda_2$, for which we assume that $\lambda = \lambda_1 = \lambda_2$.

We explore the parameter space and make a fit of the parameters using the software {\sc MultiNest} \cite{Feroz:2007kg,Feroz:2008xx,Feroz:2013hea}. In order to further improve the fit we use the software {\sc MINUIT}, and especially, the multidimensional simplex algorithm \cite{James:1975dr}. The sampling algorithm employed by {\sc MultiNest} is called nested sampling. Given the size of the parameter space, it is necessary to use such an algorithm rather than a simple parameter scan. {\sc MultiNest} is prominently designed as a tool for computation of  Bayesian evidences. In addition, as byproducts, both the posterior distribution is determined and the likelihood function, $\mathcal{L}$, is maximized. We are specifically interested in the best-fit point, which is the point that maximizes $\mathcal{L}$ or equivalently minimizes the $\chi^2$ function. The two quantities are related through
\begin{equation}
\chi^2=-2\ln{\mathcal{L}}\,.
\end{equation}
The $\chi^2$ function is defined as
\begin{equation}
\chi^2 \equiv \sum_{i=1}^N \left(\frac{X_i-\mu_i}{\sigma_i^{\rm exp}}\right)^2\,,
\end{equation}
where $X_i$ are the experimental values of the $N$ observables at $M_{\rm Z}$, $\mu_i$ are the corresponding values computed in the extended SO(10) model, and $\sigma_i^{\rm exp}$ are the experimental errors. We are not performing a Bayesian analysis, but the prior distributions needs to be specified given the Bayesian nature of {\sc MultiNest}. In our procedure, we use the priors as limits on the parameter space. For the Yukawa couplings, we ignore the scale of the couplings, and therefore, we use logarithmic priors in the interval $[10^{-12},0.1]$, where the couplings also can take on negative values. For the VEVs, we assume the prior range $[0.1,550]$, where the components of the complex VEVs also can be negative.

The procedure for the fit is the following. At the GUT scale $M_{\rm GUT}$, the values of the free parameters are generated according to the logarithmic prior distributions previously described. Furthermore, at $M_{\rm GUT}$, we fit the parameters of the Yukawa couplings and the VEVs, whereas at $M_{\rm I}$, the Higgs self-couplings. Then, we perform the RG evolution from $M_{\rm GUT}$ via $M_{\rm I}$ to $M_{\rm Z}$. We evolve the gauge couplings according to the RGEs given in Secs.~\ref{sec:RGEgaugecoupling_above} and \ref{sec:RGEgaugecoupling} and the Yukawa couplings according to the RGEs in Secs.~\ref{sec:Yukawa_above} and \ref{sec:Yukawa_below}. Next, at $M_{\rm I}$, we impose the matching conditions presented in Sec.~\ref{sec:matching}. Finally, at $M_{\rm Z}$, the observables in the extended SO(10) model are computed, and thus, a comparison to the known SM observables is performed for the quark masses (six observables), the charged-lepton masses (three observables), the quark mixing parameters (four observables), the leptonic mixing angles (three observables), and the ratio of the neutrino mass-squared differences (one observable). This procedure is repeated until the algorithm converges to a best-fit point.

The experimental values of the 17 observables are given in Tab.~\ref{tab:quantities}.
\begin{table}[ht!]
\centering
\begin{tabular}{| l | c | c | l | c | c |}
\hline\hline
\multicolumn{3}{|c|}{Quark sector } & \multicolumn{3}{|c|}{Lepton sector}\\
\hline
Observable&$X_i$&$\sigma^{\rm exp}_i$&Observable&$X_i$&$\sigma^{\rm exp}_i$\\
\hline
$m_{d}$ (GeV)& $2.90\cdot 10^{-3}$&$1.22\cdot 10^{-3}$&$m_{e}$ (GeV)&$4.87\cdot 10^{-4}$ &$2.43\cdot 10^{-5}$ \\
$m_{s}$ (GeV)& $5.50\cdot 10^{-2}$& $1.55\cdot 10^{-2}$&$m_{\mu}$ (GeV)&$1.03\cdot 10^{-1}$&$5.14\cdot 10^{-3}$\\
$m_{b}$ (GeV) &2.89& $9\cdot 10^{-2}$&$m_{\tau}$ (GeV)&$1.75$ &$8.73\cdot 10^{-2}$\\
$m_{u}$ (GeV)&$1.27\cdot 10^{-3}$&$4.6\cdot 10^{-4}$&$r \equiv \tfrac{\Delta m_{21}^2}{\Delta m_{31}^2}$ &$3\cdot 10^{-2}$&$3\cdot 10^{-3}$\\
$m_{c}$ (GeV)&$6.19\cdot 10^{-1}$&$8.4\cdot 10^{-2}$&$\sin^2\theta^\ell_{12}$&$3.06 \cdot 10^{-1}$&$1.2\cdot 10^{-2}$\\
$m_{t}$ (GeV)&$172$&$3$&$\sin^2\theta^\ell_{13}$&$2.17 \cdot 10^{-2}$&$7.5 \cdot 10^{-4}$\\
$\sin\theta^{q}_{12}$&$2.25\cdot 10^{-1}$&$1.1\cdot 10^{-3}$&$\sin^2\theta^\ell_{23}$&$4.41 \cdot 10^{-1}$&$2.7 \cdot 10^{-2}$\\
$\sin\theta^{q}_{13}$&$3.5\cdot 10^{-3}$&$3\cdot 10^{-4}$&&&\\
$\sin\theta^{q}_{23}$&$4.2\cdot 10^{-2}$&$1.3\cdot 10^{-3}$&&&\\
$\delta_{\rm CKM}\, (\pi)$& $3.88\cdot 10^{-1}$ & $1.83\cdot 10^{-2}$&&&\\
\hline
\hline
\end{tabular}
\caption{\it The 17 observables used in the $\chi^2$ function for the parameter fit at the GUT scale. The experimental values $\{X_i\}$ of the observables are the values of the observables at the EW scale and the values $\{\sigma_i^{\rm exp}\}$ are the respective experimental errors. The values of the quark and charged-lepton masses are taken from Ref.~\cite{Xing:2007fb}, the quark mixing parameters from Ref.~\cite{Joshipura:2011nn}, and the ratio of the neutrino mass-squared differences and the leptonic mixing angles from Refs.~\cite{Esteban:2016qun,NuFIT3.0}.
}\label{tab:quantities}
\end{table}
Numerically, we only fit the ratio of the neutrino mass-squared differences $r \equiv \Delta m_{21}^2/\Delta m_{31}^2$. The absolute value of the differences is determined by $\langle \overline{\Delta_R}\rangle$, see Eq.~\eqref{triplet}, which is a free parameter that only affects the size of the neutrino masses and therefore can be fitted afterwards. Furthermore, instead of using the experimental errors for the charged-lepton masses, we define the errors to be $5~\%$ of the respective masses. The experimental errors of the charged-lepton masses would render the fit virtually impossible, since they are so small that even small deviations from the experimental value would have a significant impact on the $\chi^2$ function. Finally, the VEVs are defined in Eq.~\eqref{eq:vevs} and the fit of them is carried out in such a way that $\sqrt{k_u^2+k_d^2}=246~{\rm GeV}$.

\section{Numerical Results}\label{sec:results}

We follow the procedure described in Sec.~\ref{sec:numerics} and thus perform the fit of the free parameters at $M_{\rm GUT}$ in the extended SO(10) model in such a way that we fit the observables at $M_{\rm Z}$. We find the best-fit point with $\chi^2 \simeq 11.2$ and the best-fit values of the Yukawa coupling matrices at $M_{\rm GUT}$ are the following\footnote{We present all numerical output with three significant figures.}

\begin{align}
h&\simeq { \left( {\small \begin{array}{ccc}
-2.20\cdot 10^{-6} & 0 & 0 \\
0 & -8.91\cdot 10^{-4} & 0 \\
0 & 0 & 0.294\end{array} } \right)}\,, \label{eq:h}\\
f&\simeq { \left( {\small \begin{array}{ccc}
0.00&1.78\cdot 10^{-7}-4.82\cdot 10^{-9}{\rm i} & -3.46\cdot 10^{-5}-5.84\cdot 10^{-5}{\rm i}\\
1.78\cdot 10^{-7}-4.82\cdot 10^{-9}{\rm i}  & 2.77\cdot 10^{-8}-1.34\cdot 10^{-5}{\rm i}& 1.49\cdot 10^{-4}-3.70\cdot 10^{-4}{\rm i} \\
-3.46\cdot 10^{-5}-5.84\cdot 10^{-5}{\rm i}& 1.49\cdot 10^{-4}-3.70\cdot 10^{-4}{\rm i} & 2.22\cdot 10^{-5}+2.09\cdot 10^{-4}{\rm i}\end{array} } \right)}\,, \label{eq:f}\\
g&\simeq { \left( {\small \begin{array}{ccc}
0 & 1.69\cdot 10^{-9}-4.93\cdot 10^{-7}{\rm i} & 1.62\cdot 10^{-7}+5.44\cdot 10^{-7}{\rm i}\\
-1.69\cdot 10^{-9}+4.93\cdot 10^{-7}{\rm i}  & 0 & -2.78\cdot 10^{-4}+9.51\cdot 10^{-5}{\rm i}\\
-1.62\cdot 10^{-7}-5.44\cdot 10^{-7}{\rm i}& 2.78\cdot 10^{-4}-9.51\cdot 10^{-5}{\rm i}& 0 \end{array} } \right)}\,. \label{eq:g}
\end{align}
  
For the VEVs, we first find that $r_v \simeq 55.7$~GeV, which implies that $k_u \simeq 245.9$~GeV and $k_d \simeq 4.44$~GeV. For the other VEVs, we find the following values $t_d \simeq 342$~GeV, $z_d \simeq 130$~GeV, $v_u \simeq (2.47-0.287{\rm i})$~GeV, $v_d \simeq (91.3+187{\rm i})$~GeV, $t_u \simeq (1.61-0.772{\rm i})$~GeV, and $z_u \simeq (158+2.05{\rm i})$~GeV. However, note that we have the freedom of rescaling all the VEVs and the Yukawa couplings with an overall factor.

Furthermore, we find that the best-fit value of the Higgs self-coupling $\lambda$, which is introduced at $M_{\rm I}$ with the requirement that $\lambda = \lambda_1 = \lambda_2$, is given by $\lambda \simeq 0.677$. In order to limit the number of free parameters and constraints, we only fit the ratio of the neutrino mass-squared differences $r$. The absolute value of the mass-squared differences is then determined by the value of $\langle \overline{\Delta_R}\rangle$, which, in principle, should be considered a free parameter. However, since this parameter only affects the magnitude of the neutrino masses and nothing else, it can be determined after performing the fit using the experimental value for $\Delta m^2_{21}=7.50\cdot 10^{-5}~{\rm eV^2}$. Thus, the resulting value is $\langle \overline{\Delta_R}\rangle \simeq 7.03\cdot 10^{13}$~GeV. Furthermore, we determine the values of the masses for the three neutrino mass eigenstates to be $m_1 \simeq 2.81\cdot 10^{-3}$~eV, $m_2 \simeq 9.10\cdot 10^{-3}$~eV, and $m_3 \simeq 0.0502$~eV.  Note that the fit has been performed using normal neutrino mass ordering, and in fact, inverted neutrino mass ordering cannot be accommodated.

We obtain the values of the observables in the extended SO(10) model at $M_{\rm Z}$, which are given in Tab.~\ref{tab:extendedEW} together with the values of the pulls, $g_i$, which are defined as
\begin{equation}
g_i \equiv \frac{X_i-\mu_i}{\sigma_i^{\rm exp}}\,.
\end{equation}
\begin{table}[ht!]
\centering
\begin{tabular}{| l | c | c | l | c | c |}
\hline\hline
\multicolumn{3}{|c|}{Quark sector } & \multicolumn{3}{|c|}{Lepton sector}\\
\hline
Observable&$ \mu_i $&$g_i$&Observable&$ \mu_i$&$g_i$\\
\hline
$m_{d}$ (GeV) & $1.72 \cdot 10^{-4}$ & $2.25$ &$m_{e}$ (GeV)&$4.88\cdot 10^{-4}$ &$-0.0597$ \\
$m_{s}$ (GeV) & $0.0178$& $2.40$&$m_{\mu}$ (GeV)&$0.103$&$-0.111$\\
$m_{b}$ (GeV) & $2.89$& $0.0441$&$m_{\tau}$ (GeV)&$1.75$ &$-0.0744$\\
$m_{u}$ (GeV) & $1.53\cdot 10^{-3}$&$-0.572$&$r\equiv\frac{\Delta m_{21}^2}{\Delta m_{31}^2}$ &$0.0298$&$-0.0695$\\
$m_{c}$ (GeV)&$0.620$&$-7.15\cdot 10^{-3}$&$\sin^2\theta^\ell_{12}$&0.309&$0.0484$\\
$m_{t}$ (GeV)&$172$&$-0.0197$&$\sin^2\theta^\ell_{13}$&$0.0216$&$0.0183$\\
$\sin\theta^{q}_{12}$&$0.225$&$3.11\cdot 10^{-3}$&$\sin^2\theta^\ell_{23}$&0.441&$-0.0454$\\
$\sin\theta^{q}_{13}$&$3.46 \cdot 10^{-3}$&$0.126$&&&\\
$\sin\theta^{q}_{23}$&$0.0420$&$-9.29\cdot 10^{-3}$&&&\\
$\delta_{\rm CKM}\, (\pi)$& $0.387$ & $1.27\cdot 10^{-3}$&&&\\
\hline
\hline
\end{tabular}
\caption{\it The values at the EW scale of the 17 observables, $\{\mu_i\}$, in the extended SO(10) model presented together with their respective pulls, $\{g_i\}$.} \label{tab:extendedEW}
\end{table}
We find that the observables that are clearly the most difficult to fit in this model are the quark masses $m_d$ and $m_s$. The other observables can be fitted to good accuracy.  Furthermore, at $M_{\rm Z}$, we can predict values of three unknown quantities in the lepton sector, i.e.~the leptonic Dirac CP-violating phase $\delta$  ($\delta \in [0,2\pi)$) and the two leptonic Majorana CP-violating phases $\alpha_{21}$ and $\alpha_{31}$ ($\alpha_{21},\alpha_{31} \in [0,4\pi)$) as defined in Ref.~\cite{Agashe:2014kda}. In addition, we compute three effective parameters related to the leptons: the sum of the masses for the three neutrino mass eigenstates $\Sigma$, the effective neutrino mass parameter measured in single beta decay experiments $m_{\nu_e}$, and the effective electron neutrino mass parameter that could be measured in neutrinoless double beta decay experiments $m_{ee}$, which are defined as
\begin{align}
\Sigma &\equiv m_1 +  m_2 + m_3 \,, \label{eq:Sigma}\\
m_{\nu_e} &\equiv \sqrt{m_1^2 \cos^2 \theta^\ell_{13} \cos^2 \theta^\ell_{12} + m_2^2 \cos^2 \theta^\ell_{13} \sin^2 \theta^\ell_{12} + m_3^2 \sin^2 \theta^\ell_{13}} \,, \label{eq:mnue}\\
m_{ee} &\equiv \left| \left( m_1 \cos^2 \theta^\ell_{12} + m_2 \sin^2 \theta^\ell_{12} e^{{\rm i} \alpha_{21}} \right) \cos^2 \theta^\ell_{13} + m_3 \sin^2 \theta^\ell_{13} e^{{\rm i} (\alpha_{31} - 2 \delta)} \right| \,. \label{eq:mee}
\end{align}
We find the following predicted values
\begin{align}
& \delta \simeq 0.883\pi\,, \quad \alpha_{21} \simeq 1.21\pi\,, \quad \alpha_{31} \simeq 2.55\pi\,, \nonumber\\
& \Sigma \simeq 0.0621~{\rm eV}\,, \quad  m_{\nu_e}\simeq 9.22\cdot 10^{-3}~{\rm eV}\,, \quad m_{ee} \simeq 1.53~{\rm meV}\,.
\label{eq:prediction_extended}
\end{align}     
The predicted value of $\delta$ indicates that the extended SO(10) model is about 10~\% away from being leptonic CP-conserving ($\delta = 0,\pi$) and the value of $m_{ee}$ lies perfectly within the allowed $3\sigma$ region based on a recent global fit of neutrino oscillation data \cite{Capozzi:2013csa} for normal neutrino mass ordering when $m_1 \simeq 2.81\cdot 10^{-3}$~eV \cite{DellOro:2014ysa}. Concerning the other two effective neutrino mass parameters $\Sigma$ and $m_{\nu_e}$, they lie safely below the current experimental 95~\% C.L.~upper bounds \cite{Lobashev:2003kt,Kraus:2004zw,Aseev:2011dq,Ade:2015xua}.

\begin{figure}
\begin{center}
\subfigure{\includegraphics[scale = 0.45]{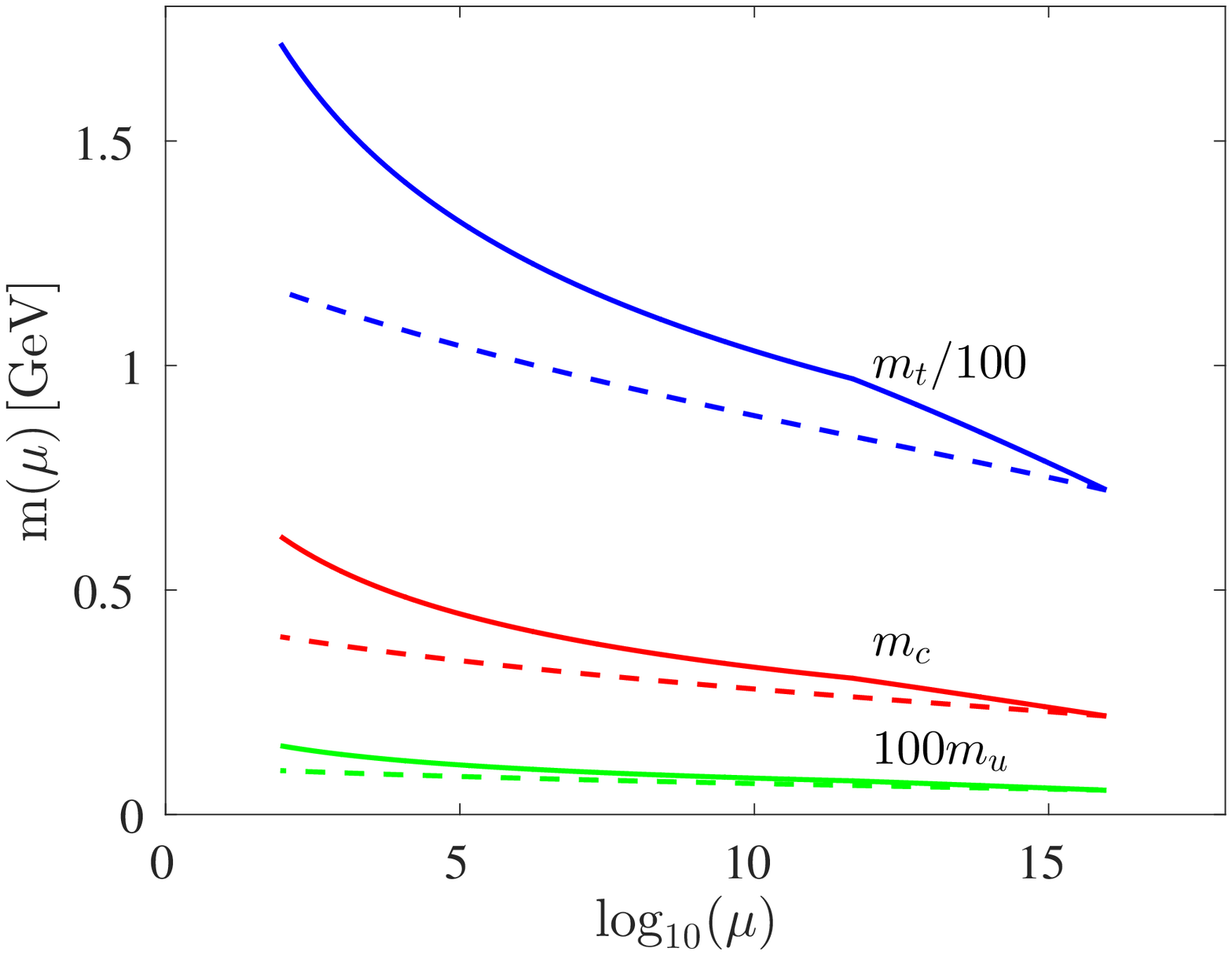}}
\subfigure{\includegraphics[scale = 0.45]{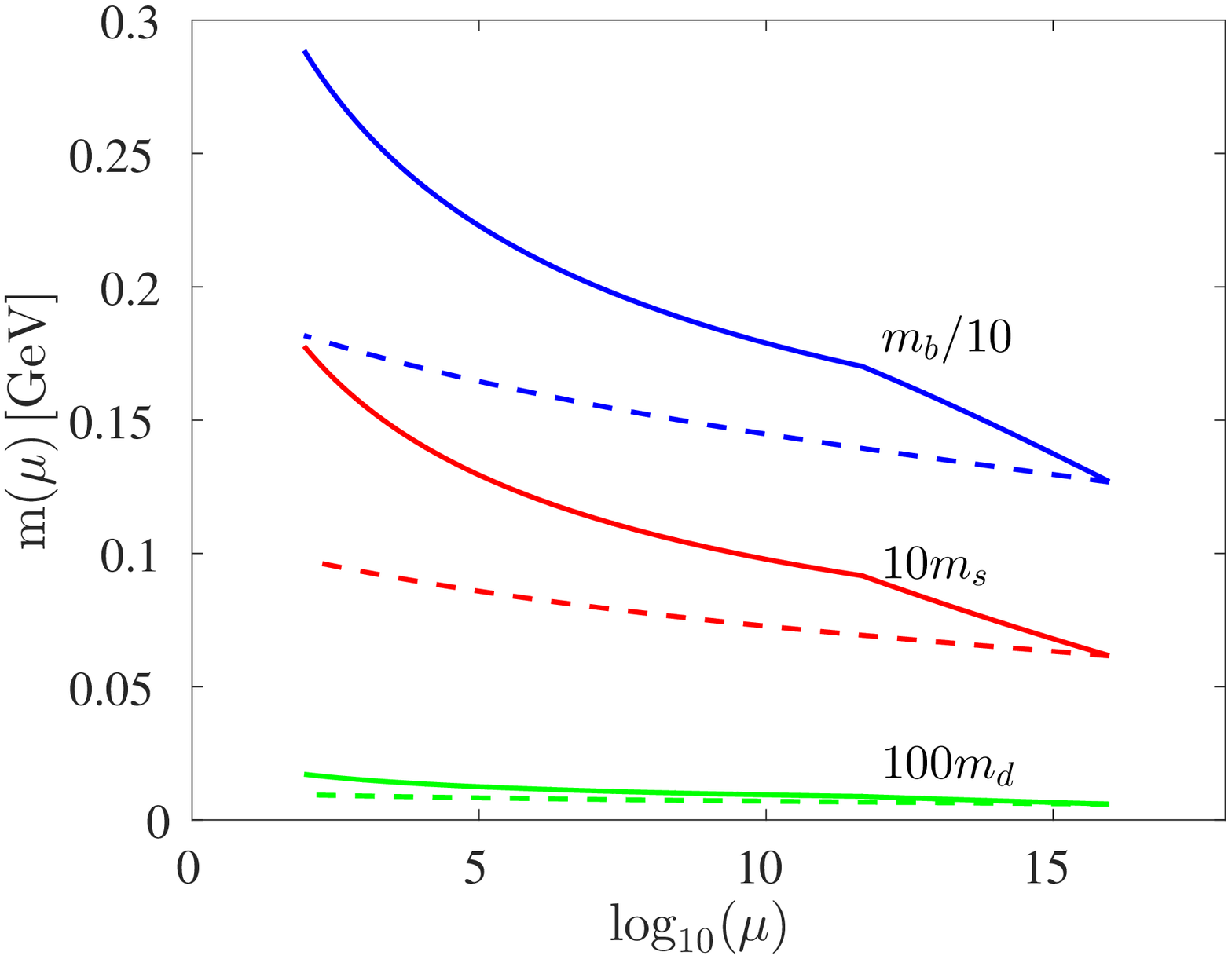}}
\caption{\it The RG running of the up-type (left panel) and down-type (right panel) quark masses, respectively, with (solid curves) and without (dashed curves) the intermediate energy scale $M_{\rm I}$ as functions of the energy scale $\mu$.}
\label{fig:quarkmasses}
\end{center}
\end{figure}

\begin{figure}
\begin{center}
\subfigure{\includegraphics[scale=0.45]{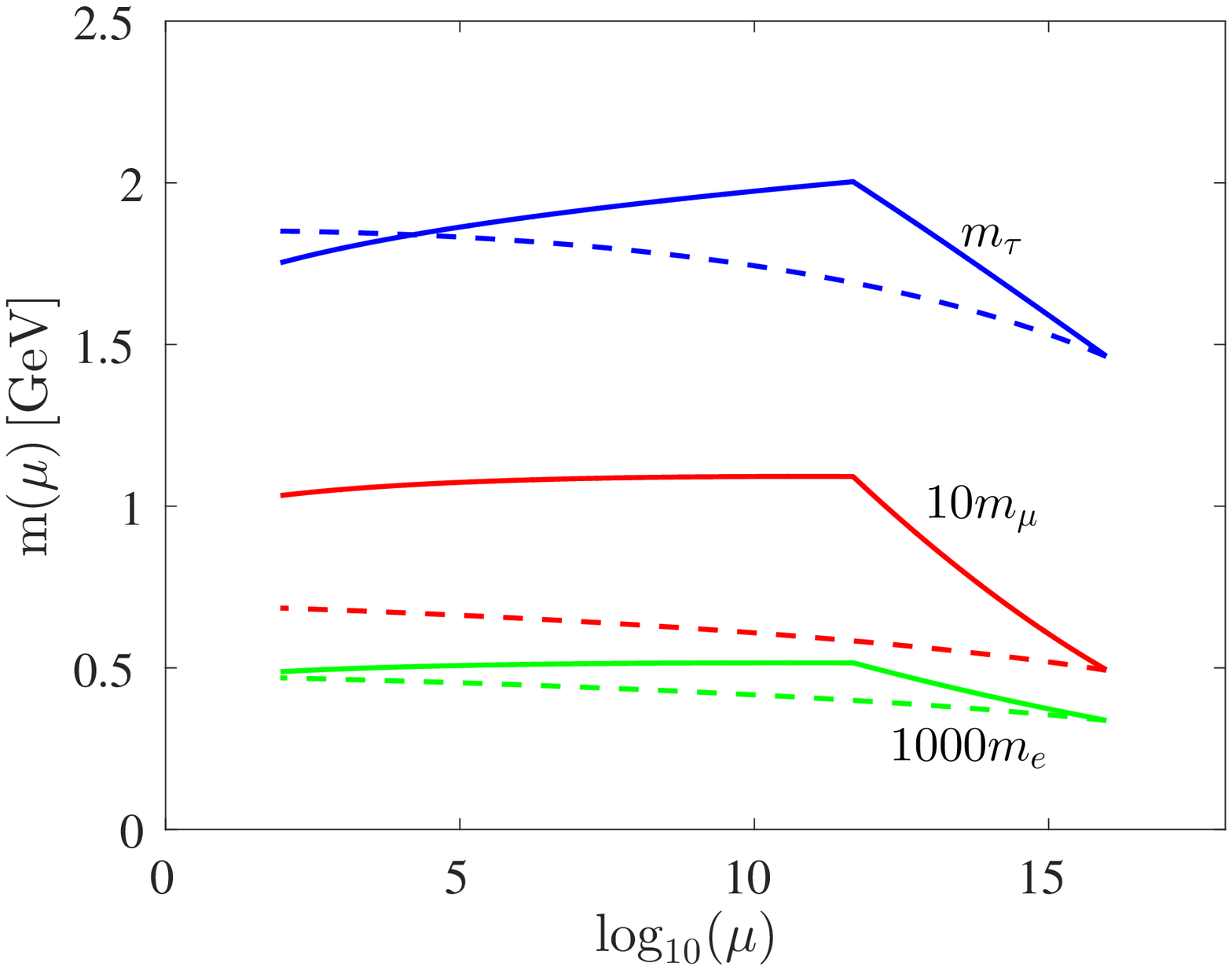}}
\subfigure{\includegraphics[scale=0.45]{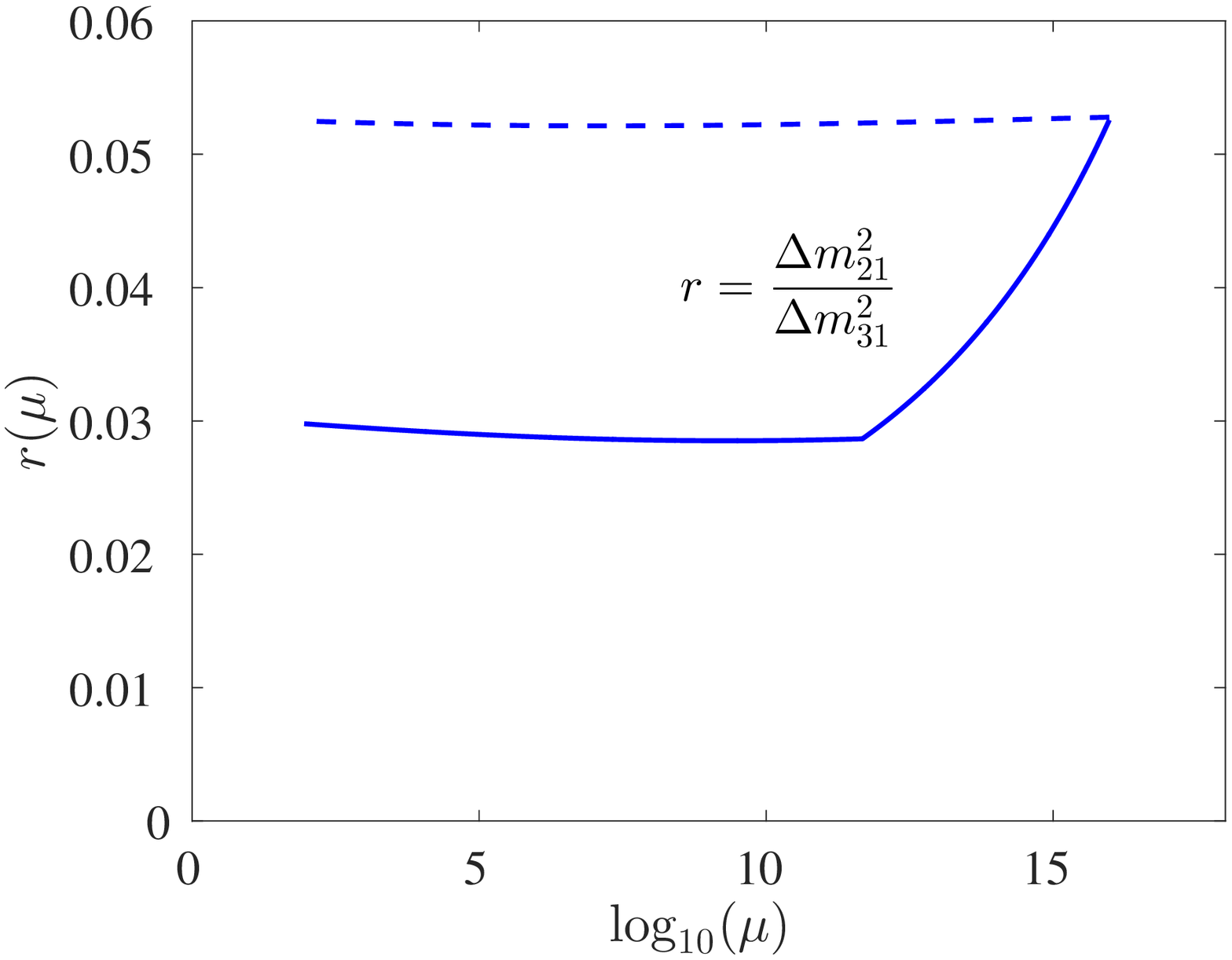}}
\caption{\it The RG running of the charged lepton masses (left panel) and the ratio of the small and large neutrino mass-squared differences (right panel), respectively, with (solid curves) and without (dashed curves) the intermediate energy scale $M_{\rm I}$ as functions of the energy scale $\mu$.}
\label{fig:leptonmasses}
\end{center}
\end{figure}

\begin{figure}
\begin{center}
\subfigure{\includegraphics[scale=0.45]{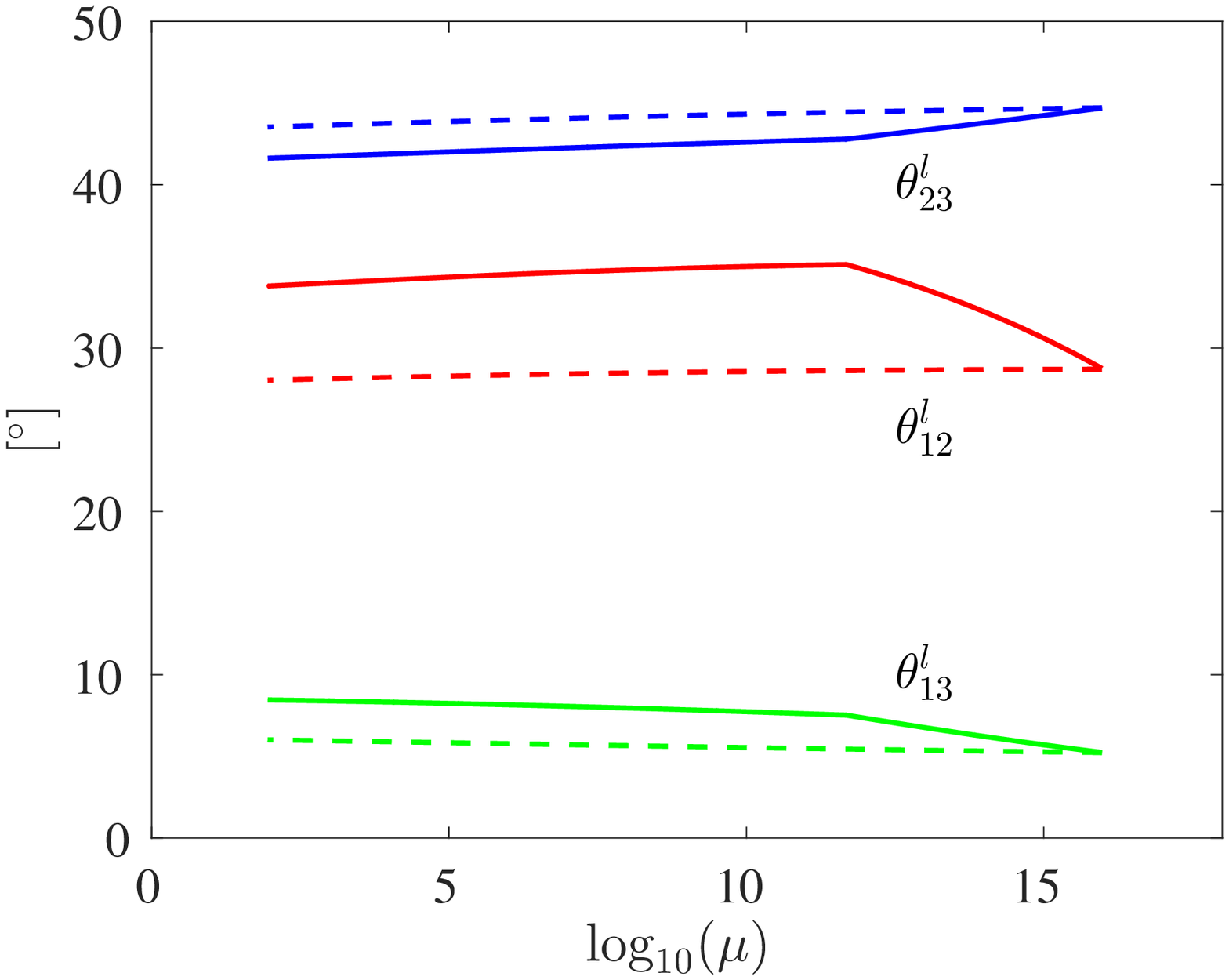}}
\subfigure{\includegraphics[scale=0.45]{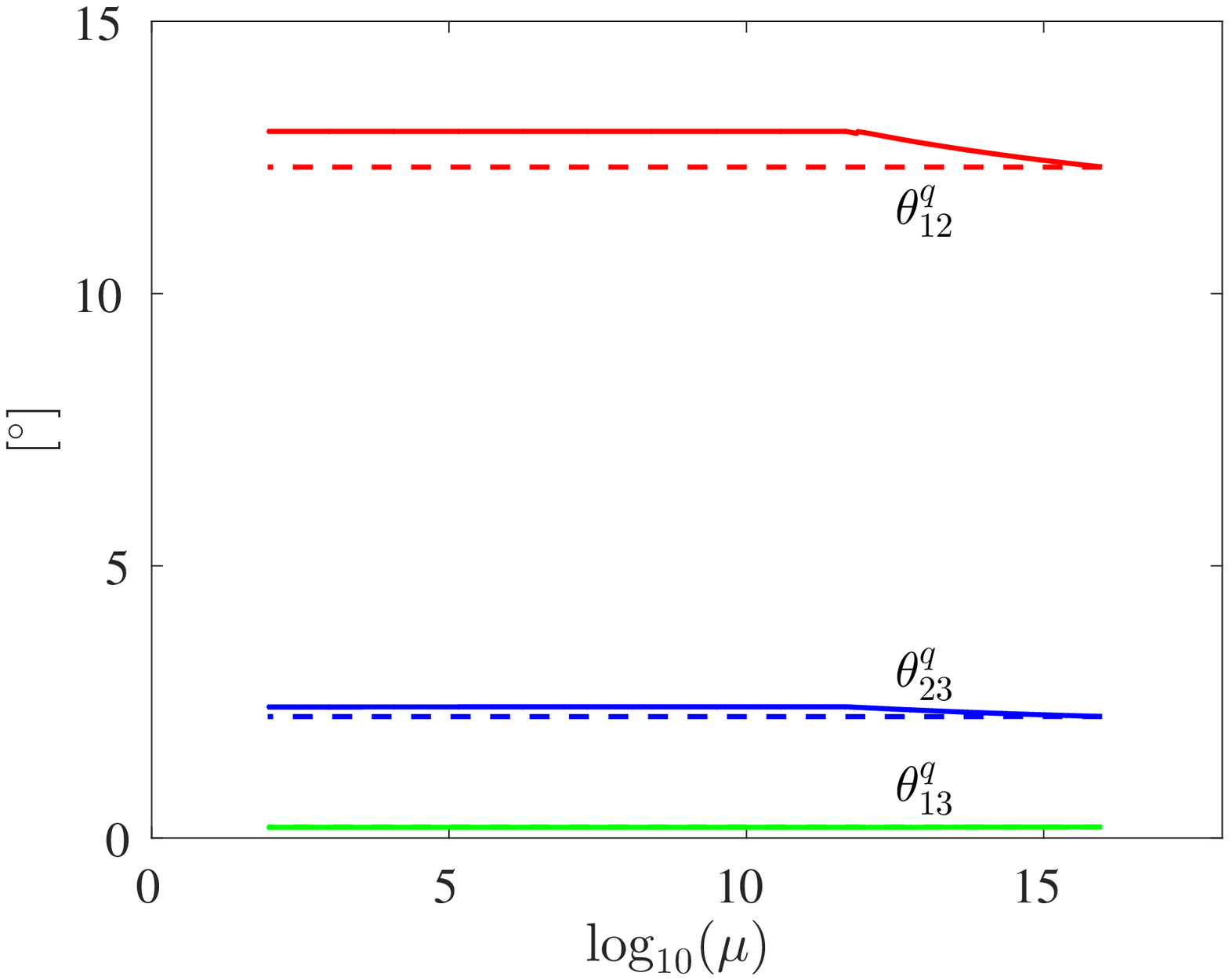}}
\caption{\it The RG running of the leptonic mixing angles (left panel) and the quark mixing angles (right panel), respectively, with (solid curves) and without (dashed curves) the intermediate energy scale $M_{\rm I}$ as functions of the energy scale $\mu$.}
\label{fig:angleRun}
\end{center}
\end{figure}

In Figs.~\ref{fig:quarkmasses}--\ref{fig:angleRun}, we present the RG running from $M_{\rm GUT}$ to $M_{\rm Z}$ for the three up-type and three down-type quark masses, the three charged-lepton masses, the three quark mixing angles, the three leptonic mixing angles, and $r$. In order to estimate the impact of the intermediate scale $M_{\rm I}$, we also present the RG running from $M_{\rm GUT}$ with the RGEs for the SM-like model with two Higgs doublets, i.e.~corresponding to the case where the intermediate scale is placed at $M_{\rm GUT}$. The RG running in this model is presented with dashed curves in the figures. We perform the comparison so that the observables in the two models coincide at $M_{\rm GUT}$. Thus, we do not fit the experimental values at $M_{\rm Z}$ in the SM-like model.

In Fig.~\ref{fig:quarkmasses}, we show the RG running of the up-type and down-type quark masses, respectively, in the extended SO(10) model with an intermediate scale (solid curves) and the SM-like model without an intermediate scale (dashed curves). For both types of masses, the slope of the RG running above $M_{\rm I}$ is larger in the extended SO(10) model than in the SM-like model, and in addition, there is a kink at $M_{\rm I}$. As a consequence, the up-type quark masses in the extended SO(10) model are larger by 45~\% -- 60~\% than in the SM-like model and the down-type quark masses by 60~\% -- 80~\%. Furthermore, for all quark masses, the direction of the RG running is the same above and below $M_{\rm I}$. In the left panel of Fig.~\ref{fig:leptonmasses}, we display the RG running of the charged-lepton masses in the extended SO(10) model (solid curves) and the SM-like model (dashed curves), whereas in the right panel of Fig.~\ref{fig:leptonmasses}, we display the RG running of $r$. Similarly, in the case of the charged-lepton masses, there is significant RG running above $M_{\rm I}$ in the extended SO(10) model, where the masses are increasing when running from $M_{\rm GUT}$ to $M_{\rm I}$. Again, there is a kink at $M_{\rm I}$, where the direction of the RG running changes, and thus, the value of the charged-lepton masses are decreasing from $M_{\rm I}$ to $M_{\rm Z}$. Hence, at $M_{\rm Z}$, $m_{e}$ and $m_{\mu}$ in the extended SO(10) model are larger by 4.0~\% and 51~\%, respectively, than in the SM-like model, whereas $m_{\tau}$ is larger in the SM-like model by 5.3~\% than in the extended SO(10) model. The main contribution to the RG running for $r$ is above $M_{\rm I}$, where the value of $r$ is decreasing from $M_{\rm GUT}$ to $M_{\rm I}$. Below $M_{\rm I}$, it is mildly increasing down to $M_{\rm Z}$, and finally, there is a significant difference for $r$ between the two models, where $r$ in the extended SO(10) model is about half the size compared to the SM-like model. In Fig.~\ref{fig:angleRun}, we present the RG running of the leptonic mixing angles in the left panel and the quark mixing angles in the right panel in the two models. For the leptonic mixing angles, the main effect on the RG running is again above $M_{\rm I}$. Both $\theta^\ell_{12}$ and $\theta^\ell_{13}$ are larger in the extended SO(10) model than in the SM-like model by 21~\% and 41~\%, respectively, whereas $\theta^\ell_{23}$ is smaller by 4.4~\%. Moreover, both $\theta^q_{12}$ and $\theta^q_{23}$ are larger by 5.3~\% and 8.0~\%, respectively, in the extended SO(10) model than in the SM-like model. The value of $\theta^q_{13}$ is 2.6~\% smaller in the extended SO(10) model than in the SM-like model.

As shown in Ref.~\cite{Mohapatra:1992dx}, threshold corrections can have a dramatic impact on $M_{\rm GUT}$ and the prediction of the proton lifetime. In our context, both $M_{\rm GUT}$ and $M_{\rm I}$ can be different from the ones computed using only the one-loop RGEs. This would in turn imply different RG running of the fermion observables. The quantification of such effects is beyond the scope of our work.

\section{Comparison between the Minimal and the Extended SO(10) Models}
\label{sec:comparison}

For comparison, we also perform a fit for the minimal non-supersymmetric SO(10) model with an intermediate scale $M_{\rm I}$. This model is presented in App.~\ref{sec:minimal}, where the RGEs of the gauge and Yukawa couplings are given. The RGEs of the Yukawa couplings in this appendix are the corrected versions of the corresponding RGEs in Refs.~\cite{Fukuyama:2002vv,Meloni:2014rga}. Thus, we repeat the procedure and update the results in Ref.~\cite{Meloni:2014rga}, although the number of Higgs fields changed from four to two at $M_{\rm Z}$. In this minimal SO(10) model, there are in total 21 free parameters. There are 15 free parameters in the Yukawa coupling matrices, defined in Eq.~\eqref{eq:yuk}, which are three real parameters in the real diagonal matrix $h$ and six complex parameters in the symmetric matrix $f$. In addition, there are five parameters related to the VEVs, $r_v=k_u/k_d$ (one real parameter) and $v_u$ and $v_d$ (two complex parameters). Finally, we fit the Higgs self-coupling constant $\lambda = \lambda_1 = \lambda_2$, which is introduced at $M_{\rm I}$. Similar to the case of the extended SO(10) model, we assume in the minimal SO(10) model that there are only two Higgs doublets below $M_{\rm I}$. Furthermore, we use the same numerical parameter-fitting procedure as discussed in Sec.~\ref{sec:numerics} for the minimal SO(10) model.

In the minimal SO(10) model, the $\chi^2$ function for the best-fit point is $\chi^2 \simeq 8.93$. Thus, we can fit the observables at $M_{\rm Z}$ in the minimal SO(10) model as well. The best-fit values for the Yukawa coupling matrices at $M_{\rm GUT}$ are given by
\begin{align}
h&\simeq { \left( {\small \begin{array}{ccc}
2.21\cdot 10^{-6} & 0 & 0 \\
0 & -1.65\cdot 10^{-3} & 0 \\
0 & 0 & -0.508\end{array} } \right)}\,,\label{eq:h_min}\\
f&\simeq { \left( {\small \begin{array}{ccc}
3.99\cdot 10^{-6}-2.31\cdot 10^{-5}{\rm i} &5.74\cdot 10^{-6}+1.32\cdot 10^{-4}{\rm i} & -1.55\cdot 10^{-2}-4.10\cdot 10^{-2}{\rm i}\\
5.74\cdot 10^{-6}+1.32\cdot 10^{-4}{\rm i}  & 8.08\cdot 10^{-7}+4.59\cdot 10^{-4}{\rm i}& -0.154+6.25\cdot 10^{-5}{\rm i} \\
 -1.55\cdot 10^{-2}-4.10\cdot 10^{-2}{\rm i}& -0.154+6.25\cdot 10^{-5}{\rm i} & -6.89\cdot 10^{-2}-7.58\cdot 10^{-5}{\rm i}\end{array} } \right)}\,. \label{eq:f_min}
\end{align}
For the parameters related to the VEVs, we find $r_v \simeq 55.1$~GeV, $v_u \simeq 7.95$~GeV, and $v_d \simeq (0.512+1.01{\rm i})$~GeV, whereas we find the Higgs self-coupling to be $\lambda \simeq 2.88\cdot 10^{-2}$. Furthermore, for the neutrino masses, we find that $\langle \overline{\Delta_R}\rangle \simeq 3.46\cdot 10^{12}$~GeV, and therefore, we determine the values of the masses for the three neutrino mass eigenstates to be $m_1 \simeq 1.43\cdot 10^{-3}$~eV, $m_2 \simeq 8.78\cdot 10^{-3}$~eV, and $m_3 \simeq 0.0505$~eV. Again, the fit has been performed using normal neutrino mass ordering. Finally, at $M_{\rm Z}$, we obtain the values of the observables, $\mu_i$, together with their corresponding pulls, $g_i$, in the minimal SO(10) model, which are given in Tab.~\ref{tab:minimalEW}. 
\begin{table}[ht!]
\centering
\begin{tabular}{| l | c | c | l | c | c |}
\hline\hline
\multicolumn{3}{|c|}{Quark sector } & \multicolumn{3}{|c|}{Lepton sector}\\
\hline
Observable&$ \mu_i $&$g_i$&Observable&$ \mu_i$&$g_i$\\
\hline
$m_{d}$ (GeV) & $1.16 \cdot 10^{-3}$ & $1.43$ &$m_{e}$ (GeV)&$4.87\cdot 10^{-4}$ &$-0.0165$ \\
$m_{s}$ (GeV) & $0.0326$& $1.44$&$m_{\mu}$ (GeV)&$0.103$&$-0.0928$\\
$m_{b}$ (GeV) & $2.98$& $-0.961$&$m_{\tau}$ (GeV)&$1.61$ &$1.61$\\
$m_{u}$ (GeV) & $1.33\cdot 10^{-3}$&$-0.138$&$r\equiv\frac{\Delta m_{21}^2}{\Delta m_{31}^2}$ &$0.0294$&$-0.188$\\
$m_{c}$ (GeV)&$0.704$&$-1.02$&$\sin^2\theta^\ell_{12}$&0.305&$0.0899$\\
$m_{t}$ (GeV)&$171$&$0.337$&$\sin^2\theta^\ell_{13}$&$0.0217$&$-0.0676$\\
$\sin\theta^{q}_{12}$&$0.225$&$0.0549$&$\sin^2\theta^\ell_{23}$&0.443&$-0.0684$\\
$\sin\theta^{q}_{13}$&$3.55 \cdot 10^{-3}$&$-0.177$&&&\\
$\sin\theta^{q}_{23}$&$0.0421$&$-0.0696$&&&\\
$\delta_{\rm CKM}\, (\pi)$& $0.388$ & $-0.0313$&&&\\
\hline
\hline
\end{tabular}
\caption{\it The values at the EW scale of the 17 observables, $\{\mu_i\}$, in the minimal SO(10) model presented together with their respective pulls, $\{g_i\}$.} \label{tab:minimalEW}
\end{table}
 
We find that the observables that are the most difficult to fit in this model are the following masses: $m_d$, $m_s$, $m_c$, $m_b$, and $m_{\tau}$. This is similar to the result in the extended SO(10) model, and furthermore, we can conclude that masses seems to be more difficult to fit compared to the mixing parameters. Similarly, as for the extended SO(10) model, we are able to compute predicted values at $M_{\rm Z}$ in this model for the three unknown quantities in the lepton sector as well as the three effective neutrino mass parameters using Eqs.~\eqref{eq:Sigma}--\eqref{eq:mee}, which are 
\begin{align}
& \delta \simeq 0.426\pi\,, \quad \alpha_{21} \simeq 1.02\pi\,, \quad \alpha_{31} \simeq 2.07\pi\,, \nonumber\\
& \Sigma \simeq 0.0607~{\rm eV}\,, \quad m_{\nu_e} \simeq 0.00893~{\rm eV}\,, \quad m_{ee} \simeq 2.64~{\rm meV}\,.
\end{align}
where $\delta \sim \pi/2$ means that the model exhibits maximal leptonic CP violation. As for the extended model (cf.~Eq.~\eqref{eq:prediction_extended}), the predicted value of $\delta$ for the minimal SO(10) model is not similar to the best-fit value from the latest global fit to neutrino oscillation data, which is around $3\pi/2$ \cite{Esteban:2016qun}. Nevertheless, all three values of the effective neutrino mass parameters are below the current experimental 95~\% C.L.~upper bounds \cite{Lobashev:2003kt,Kraus:2004zw,Aseev:2011dq,Ade:2015xua}.

\section{Summary and Conclusions}
\label{summary}

We have investigated the RG evolution of fermion observables (i.e.~the fermion masses and mixings) in an extended non-supersymmetric SO(10) model with the PS group as the intermediate group including the ${\bf 10}_{\rm H}$, ${\bf 126}_{\rm H}$, and ${\bf 120}_{\rm H}$ Higgs representations. We have determined the RGEs for the gauge and Yukawa couplings, and most importantly, the ones for the Yukawa couplings in the range between the GUT scale $M_{\rm GUT}$ and the intermediate scale $M_{\rm I}$. At $M_{\rm GUT}$, we have introduced the free parameters in the Yukawa couplings in $h$, $f$, and $g$ including parameters related to the VEVs. Below $M_{\rm I}$, we have assumed a SM-like model with two Higgs doublets. We have performed the RG evolution of the fermion observables and numerically computed a fit in the extended SO(10) model. We have found that all the observables can be satisfyingly fitted at the EW scale for normal neutrino mass ordering. Moreover, we have found that a satisfactory fit is not possible to obtain for inverted neutrino mass ordering. The RG running is significant for all the parameters, although especially so for the fermion masses and the ratio of the neutrino mass-squared differences. The slope of the RG running, and sometimes also the sign of its derivative, is changed at $M_{\rm I}$. In addition, we have made a comparison of this model to a SM-like model without the intermediate scale. Since the intermediate scale has a significant effect on the results at the EW scale, the values of the quark masses in the two models differ by 45~\% -- 80~\%. The differences of the values of the lepton masses are smaller, especially $m_\tau$ will be smaller in the SM-like model. The RG running for the mixings is less significant than the one for the fermion masses, and hence, the difference in the values of mixings at the EW scale is smaller than the corresponding difference for the fermion masses. Furthermore, we have presented predictions for the values of the masses of the three neutrino mass eigenstates, the three Dirac and Majorana CP-violating phases, and three effective neutrino masses. In particular, we have found that the value of the Dirac CP-violating phase is about 10~\% away from $\pi$, which means that the extended SO(10) model is nearly CP-conserving. Finally, we have also made a fit in a more minimal SO(10) model with an intermediate scale, where the Higgs representations are ${\bf 10}_{\rm H}$ and ${\bf 126}_{\rm H}$. For this model, we have again assumed that there are two Higgs doublets below the intermediate scale. We have found that the fermion observables can be well accommodated at the EW scale for normal neutrino mass ordering in this minimal model too.

\section*{Acknowledgments}

We would like to thank K.S.~Babu and S.~Khan for valuable discussions in the initial stage of this work. We would also like to thank J.~Schwichtenberg for useful discussions. In addition, SR would like to thank K.~Lindvall for useful discussions. Numerical computations were performed on resources provided by the Swedish National Infrastructure for Computing (SNIC) at PDC Center for High Performance Computing (PDC-HPC) at KTH Royal Institute of Technology in Stockholm, Sweden under project number PDC-2016-76. This work was supported by MIUR (Italy) under the program Futuro in Ricerca 2010 (RBFR10O36O) (D.M.) and the Swedish Research Council (Vetenskapsr{\aa}det), contract no. 621-2011-3985 (T.O.).

\appendix

\section{Renormalization Group Equations of the Minimal SO(10) Model}
\label{sec:minimal}

In this appendix, the minimal non-supersymmetric SO(10) model is presented including the RGEs and matching conditions of this model. We evolve the RGEs to leading order in perturbation theory from the scale of a grand unified theory (GUT) $M_{\rm GUT}$ via an intermediate scale $M_{\rm I}$ down to the EW scale $M_{\rm Z} = (91.1876 \pm 0.0021)$~GeV \cite{Agashe:2014kda}. Here, we again use $M_{\rm GUT} = (1.0 \pm 0.5) \cdot 10^{16}$~GeV and $M_{\rm I} = (4.8 \pm 2.2) \cdot 10^{11}$~GeV.

First, in App.~\ref{sec:matchGUT}, we present the Yukawa sector of the Lagrangian for the minimal SO(10) model above and below $M_{\rm GUT}$ as well as the matching conditions for the Yukawa and gauge couplings at $M_{\rm GUT}$. Second, in App.~\ref{sec:RGEsGUTI}, we give the RGEs for the gauge and Yukawa couplings between $M_{\rm GUT}$ and $M_{\rm I}$. Third, in App.~\ref{sec:matchI}, we display the matching conditions for the gauge and Yukawa couplings as well as the quartic scalar Higgs self-coupling and the effective neutrino mass matrix at $M_{\rm I}$. Fourth, in App.~\ref{sec:RGEsIZ}, we give the RGEs for the gauge and Yukawa couplings, the quartic scalar Higgs self-coupling, and the effective neutrino mass matrix between $M_{\rm I}$ and $M_{\rm Z}$. Finally, in App.~\ref{sec:matchZ}, we show the matching conditions for the gauge couplings at $M_{\rm Z}$.

\subsection{The Yukawa Lagrangian of the Minimal SO(10) Model and Matching Conditions at $M_{\rm GUT}$}
\label{sec:matchGUT}

Above $M_{\rm GUT}$, the Yukawa sector of the Lagrangian is given by
\begin{equation}
- {\mathcal L}_Y^{\rm GUT} = {\bf 16} \,(h\, {\bf 10}_{\rm H}+f\, \overline{{\bf 126}}_{\rm H})\, {\bf 16} \,,
\label{eq:yuk}
\end{equation}
where $h \equiv Y^{(10)}$ and $f \equiv Y^{(126)}$ are the Yukawa couplings. Equivalently, below $M_{\rm GUT}$, the Yukawa sector of the Lagrangian for the minimal SO(10) model is given by \cite{Fukuyama:2002vv}
\begin{equation}
- {\mathcal L}_Y^{10 + \overline{126}} = Y_F^{(10)} \bar F_L \Phi F_R + Y_F^{(126)} \bar F_L \Sigma F_R + Y_R^{(126)} F_R^{\rm T} C F_R \overline{\Delta_R} \,,
\label{eq:y2}
\end{equation}
where $Y_F^{(10)}$ and $Y_F^{(126)}$ are the Yukawa couplings of the SU(4) singlet and the SU(4) 15-plet, respectively, $Y_R^{(126)}$ is the right-handed Majorana neutrino coupling, and $F_L = ({\bf 4},{\bf 2},{\bf 1})$, $F_R = ({\bf 4},{\bf 1},{\bf 2})$, $\Phi = ({\bf 1},{\bf 2},{\bf 2})$, $\Sigma = ({\bf 15},{\bf 2},{\bf 2})$, and $\overline{\Delta_R} = (\overline{{\bf 10}},{\bf 1},{\bf 3})$. Therefore, we need to match the two different sets of Yukawa couplings at $M_{\rm GUT}$. These matching conditions are \cite{Aulakh:2002zr,Fukuyama:2002vv,Meloni:2014rga}
\begin{align}
\frac{1}{\sqrt{2}}Y^{(10)}_F (M_{\rm GUT}) &\equiv Y^{(10)}(M_{\rm GUT})\,, \label{eq:bc1_min} \\
\frac{1}{4\sqrt{2}}Y^{(126)}_F(M_{\rm GUT}) =\frac{1}{4}Y^{(126)}_R (M_{\rm GUT})
&\equiv Y^{(126)}(M_{\rm GUT})\,. \label{eq:bc2}
\end{align}
In addition, note that the matching conditions for the gauge coupling constants at $M_{\rm GUT}$ are the same for the minimal SO(10) model as the ones for the extended SO(10) model described in Sec.~\ref{sec:matchingGaugeGUT}, see Eq.~\eqref{eq:matchingGaugeGUT}.
However, in this model, the value of the gauge coupling at the GUT scale is given by
\begin{equation}
\alpha_{\rm GUT} = 0.027 \,.
\end{equation}

\subsection{RGEs from $M_{\rm GUT}$ to $M_{\rm I}$}
\label{sec:RGEsGUTI}

\subsubsection{RGEs for the Gauge Couplings}

Between $M_{\rm GUT}$ and $M_{\rm I}$, the RGEs for the gauge coupling constants $g_{2L}$, $g_{2R}$, and $g_{4C}$ read
\begin{align}
16 \pi^2 \frac{{\rm d} g_{2L}}{{\rm d} t} & =  2 g_{2L}^3 \,, \label{eq:2L_min} \\
16 \pi^2 \frac{{\rm d} g_{2R}}{{\rm d} t} & = \frac{26}{3} g_{2R}^3 \,, \label{eq:2R_min} \\
16 \pi^2 \frac{{\rm d} g_{4C}}{{\rm d} t} & = - \frac{7}{3} g_{4C}^ 3 \,. \label{eq:4C_min}
\end{align}

\subsubsection{RGEs for the Yukawa Couplings}

Furthermore, between $M_{\rm GUT}$ and $M_{\rm I}$, the RGEs for the Yukawa couplings $Y_F^{(10)}$, $Y_F^{(126)}$, and $Y_R^{(126)}$ are found to be
{\allowdisplaybreaks  \begin{align} 
16 \pi^2 \frac{{\rm d} Y_F^{(10)}}{{\rm d} t} & =  
\Big({Y_F^{(10)}}  {Y_F^{(10)}}^{\dagger} + \frac{15}{4} {{Y_F^{(126)}}  {Y_F^{(126)}}^{\dagger} \Big) {Y_F^{(10)}}} \nonumber\\
&+ {Y_F^{(10)}} \Big\{ {Y_F^{(10)}}^{\dagger}  {Y_F^{(10)}} + \frac{15}{4} \Big({Y_F^{(126)}}^{\dagger}  {Y_F^{(126)}} + 2 {Y_R^{(126)}}^*  {Y_R^{(126)}}\Big)\Big\}\nonumber\\ & + 4 \, \mbox{tr}\Big({{Y_F^{(10)}}  {Y_F^{(10)}}^{\dagger}}\Big) {Y_F^{(10)}} - \frac{9}{4} \Big(g_{2L}^{2} + g_{2R}^{2} + 5 g_{4C}^{2}\Big) {Y_F^{(10)}} \,,\label{eq:YF10_min}\\ 
16 \pi^2 \frac{{\rm d} Y_F^{(126)}}{{\rm d} t} & =  
\Big({{Y_F^{(10)}}}  {Y_F^{(10)}}^{\dagger} + \frac{15}{4} {{Y_F^{(126)}}  {Y_F^{(126)}}^{\dagger} \Big) {Y_F^{(126)}}} \nonumber\\
&+ {Y_F^{(126)}} \Big\{ {Y_F^{(10)}}^{\dagger}  {Y_F^{(10)}} + \frac{15}{4} \Big({Y_F^{(126)}}^{\dagger}  {Y_F^{(126)}}  + 2 {Y_R^{(126)}}^*  {Y_R^{(126)}}\Big)\Big\}\nonumber\\ & + \mbox{tr}\Big({{Y_F^{(126)}}  {Y_F^{(126)}}^{\dagger}}\Big) {Y_F^{(126)}} - \frac{9}{4} \Big(g_{2L}^{2} + g_{2R}^{2} + 5 g_{4C}^{2}\Big) {Y_F^{(126)}} \,,\label{eq:YF126_min}\\ 
16 \pi^2 \frac{{\rm d} Y_R^{(126)}}{{\rm d} t} & =  
\Big\{{Y_F^{(10)}}^{T}  {Y_F^{(10)}}^* + \frac{15}{4} \Big({{Y_F^{(126)}}^{T}  {Y_F^{(126)}}^* + 2 {{Y_R^{(126)}}}  {Y_R^{(126)}}^*\Big)\Big\} {Y_R^{(126)}}}\nonumber\\ & + {{Y_R^{(126)}}}  \Big\{{Y_F^{(10)}}^{\dagger}  {Y_F^{(10)}} + \frac{15}{4} \Big({Y_F^{(126)}}^{\dagger}  {Y_F^{(126)}} + 2 {Y_R^{(126)}}^* {Y_R^{(126)}}\Big)\Big\}\nonumber\\ & + 2 \, \mbox{tr}\Big({{Y_R^{(126)}}  {Y_R^{(126)}}^*}\Big) {Y_R^{(126)}} - \frac{9}{4} \Big(2 g_{2R}^{2}  + 5 g_{4C}^{2}\Big) {Y_R^{(126)}} \,. \label{eq:YR126_min}
\end{align}}
Note that Eqs.~\eqref{eq:2L_min}--\eqref{eq:4C_min} and \eqref{eq:YF10_min}--\eqref{eq:YR126_min} have been computed using the software {\sc SARAH 4} \cite{Staub:2013tta}. Importantly, Eqs.~\eqref{eq:YF10_min}--\eqref{eq:YR126_min} replace Eqs.~(24)--(26) in Ref.~\cite{Fukuyama:2002vv} and Eqs.~(A.1)--(A.3) in Ref.~\cite{Meloni:2014rga}, which are not correct.

\subsection{Matching Conditions at $M_{\rm I}$}
\label{sec:matchI}

\subsubsection{Matching Conditions for the Gauge Couplings}

Note that the matching conditions for the gauge coupling constants at $M_{\rm I}$ are the same for the minimal SO(10) model as the ones for the extended SO(10) model described in Sec.~\ref{sec:matchingGauge}. 

\subsubsection{Matching Conditions for the Yukawa Couplings}

Below $M_{\rm I}$, the Yukawa sector of the Lagrangian for the two-Higgs doublet model is again given by Eq.~\eqref{eq:y}. However, at $M_{\rm I}$, the matching conditions for the different Yukawa couplings above and below $M_{\rm I}$ are chosen as
\begin{align}
Y_u (M_{\rm I}) &\equiv Y^{(10)}_F(M_{\rm I}) + \frac{v_u}{4 k_u} Y^{(126)}_F(M_{\rm I}) \,, \label{eq:matchmi1}\\
Y_d(M_{\rm I}) &\equiv Y^{(10)}_F(M_{\rm I}) + \frac{v_d}{4 k_d} Y^{(126)}_F(M_{\rm I}) \,, \label{eq:matchmi2}\\
Y_e(M_{\rm I}) &\equiv Y^{(10)}_F(M_{\rm I}) - \frac{3 v_d}{4 k_d} Y^{(126)}_F(M_{\rm I}) \,,
\label{eq:matchmi3}
\end{align}
where $k_{u,d} \equiv \langle \Phi_{u,d} \rangle_{10}$ and $v_{u,d} \equiv \langle \Sigma_{u,d} \rangle_{126}$ are VEVs of Higgs submultiplets.

\subsubsection{Matching Conditions for the Quartic Scalar Higgs Self-Couplings}

Note that we assume again the same Higgs potential below $M_{\rm I}$ as for the extended SO(10) model. In addition, the matching conditions for the quartic scalar Higgs self-couplings at $M_{\rm I}$ are the same for the minimal SO(10) model as the ones for the extended SO(10) model described in Sec.~\ref{sec:matchQSHSC}. 

\subsubsection{Matching Condition for the Effective Neutrino Mass Matrix}

We have to make a matching for the effective neutrino mass matrix at $M_{\rm I}$. Below $M_{\rm I}$, the effective neutrino mass matrix is given by Eq.~\eqref{eq:mkappa}. At $M_{\rm I}$, this expression must match the form of $m_\nu$ valid from $M_{\rm GUT}$ to $M_{\rm I}$, which can be conveniently expressed as
\begin{equation}
m_\nu = \frac{k_u^2}{2} {Y_F^{(10)}}^T M_R^{-1} Y_F^{(10)} - \frac{3}{8} k_u v_u 
\left\{ {Y_F^{(126)}}^T M_R^{-1} Y_F^{(10)} + {Y_F^{(10)}}^T M_R^{-1} Y_F^{(126)} \right\} + \frac{9}{32} v_u^2 {Y_F^{(126)}}^T M_R^{-1} Y_F^{(126)} \,,
\end{equation}
where $M_R$ is the RG running quantity defined in Eq.~\eqref{triplet}. Therefore, at $M_{\rm I}$, we have the following matching condition
\begin{align}
\kappa(M_{\rm I}) &\equiv {Y_F^{(10)}}^T(M_{\rm I}) \, M_R^{-1}(M_{\rm I}) \, Y_F^{(10)}(M_{\rm I}) \nonumber\\
&- \frac{3}{4} \frac{v_u}{k_u} 
\left\{ {Y_F^{(126)}}^T(M_{\rm I}) \, M_R^{-1}(M_{\rm I}) \, Y_F^{(10)}(M_{\rm I}) + {Y_F^{(10)}}^T(M_{\rm I}) \, M_R^{-1}(M_{\rm I}) \, Y_F^{(126)}(M_{\rm I}) \right\} \nonumber\\
&+ \frac{9}{16} \frac{v_u^2}{k_u^2} {Y_F^{(126)}}^T(M_{\rm I}) \, M_R^{-1}(M_{\rm I}) \, Y_F^{(126)}(M_{\rm I}) \,.
\end{align}

\subsection{RGEs from $M_{\rm I}$ to $M_{\rm Z}$}
\label{sec:RGEsIZ}

\subsubsection{RGEs for the Gauge Couplings}

Note that the RGEs for the gauge coupling constants from $M_{\rm I}$ to $M_{\rm Z}$ are the same for the minimal SO(10) model as the ones for the extended SO(10) model described in Sec.~\ref{sec:RGEgaugecoupling}. 

\subsubsection{RGEs for the Yukawa Couplings}

Note that the RGEs for the Yukawa couplings from $M_{\rm I}$ to $M_{\rm Z}$ are the same for the minimal SO(10) model as the ones for the extended SO(10) model described in Sec.~\ref{sec:Yukawa_below}.

\subsubsection{RGEs for the Quartic Scalar Higgs Self-Couplings}

Note that the RGEs for the quartic scalar Higgs self-couplings from $M_{\rm I}$ to $M_{\rm Z}$ are the same for the minimal SO(10) model as the ones for the extended SO(10) model described in Sec.~\ref{sec:RGEsQSHSC}.

\subsubsection{RGE for the Effective Neutrino Mass Matrix}

Note that the RGE for the effective neutrino mass matrix from $M_{\rm I}$ to $M_{\rm Z}$ is the same for the minimal SO(10) model as the one for the extended SO(10) model described in Sec.~\ref{sec:RGEsENMM}.

\subsection{Matching Conditions at $M_{\rm Z}$}
\label{sec:matchZ}

Note that the matching conditions for the gauge coupling constants at $M_{\rm Z}$ are the same for the minimal SO(10) model as the ones for the extended SO(10) model described in Sec.~\ref{sec:matchMZ}.

\providecommand{\href}[2]{#2}\begingroup\raggedright\endgroup


\end{document}